\documentclass[twocolumn,showpacs,superscriptaddress,nofootinbib]{revtex4}
\usepackage{graphicx}
\usepackage{amsmath}
\usepackage{hyperref}
\usepackage{baybicep}
\usepackage{eivars}
\usepackage{xspace}
\usepackage{longtable}
\usepackage{color}

\hypersetup{
    colorlinks=true,       
    linkcolor=red,          
    citecolor=cyan,        
    filecolor=magenta,      
    urlcolor=blue,           
}

\newcommand{\ie}{i.e.\xspace}

\newlength{\wsingfig}
\newlength{\wdblefig}
\newlength{\wfull}
\newlength{\hfull}
\newcommand{\sss}[1]{{\scriptscriptstyle{#1}}}

\newcommand{\sr}{\mathrm{SR}}
\newcommand{\up}{\mathrm{p}}
\newcommand{\uk}{\mathrm{k}}
\newcommand{\uuc}{\mathrm{uc}}
\newcommand{\uSZ}{\mathrm{SZ}}

\newcommand{\usssA}{\sss{\mathrm{A}}}
\newcommand{\usssB}{\sss{\mathrm{B}}}
\newcommand{\usssMC}{\sss{\mathrm{MC}}}
\newcommand{\usssPS}{\sss{\mathrm{PS}}}
\newcommand{\usssREF}{\sss{\mathrm{REF}}}
\newcommand{\usssSR}{\sss{\mathrm{SR}}}
\newcommand{\usssCIB}{\sss{\mathrm{CIB}}}
\newcommand{\umod}{\sss{\mathrm{mod}}}
\newcommand{\utSZ}{\ut\sss{\uSZ}}
\newcommand{\ukSZ}{\uk\sss{\uSZ}}
\newcommand{\calH}{\mathcal{H}}
\newcommand{\calP}{\mathcal{P}}

\newcommand{\calO}{\mathcal{O}}

\newcommand{\calL}{\mathcal{L}}
\newcommand{\calLmax}{\mathcal{L}^{\max}}
\newcommand{\calLp}{\mathcal{L}_{\up}}
\newcommand{\calLpbar}{\bar{\mathcal{L}}_{\up}}
\newcommand{\calLb}{\mathcal{L}_{\ub}}

\newcommand{\calR}{\mathcal{R}}
\newcommand{\Rfac}{\calR}
\newcommand{\calRref}{\mathcal{R}_{\usssREF}}
\newcommand{\calE}{\mathcal{E}}
\newcommand{\calM}{\mathcal{M}}
\newcommand{\calMref}{\mathcal{M}_{\usssREF}}
\newcommand{\calRsr}{\calR_{\sss{\sr}}}

\newcommand{\vareps}{\varepsilon}
\newcommand{\varepsfix}{\varepsilon_{\mathrm{f}}}

\newcommand{\uq}{\mathrm{q}}
\newcommand{\ucl}{\mathrm{cl}}
\newcommand{\ugw}{\mathrm{gw}}

\newcommand{\DA}{D_\usssA}
\newcommand{\DB}{D_\usssB}
\newcommand{\Db}{D_\ub}
\newcommand{\Dp}{D_\up}

\newcommand{\thetaA}{\theta_\usssA}
\newcommand{\thetaB}{\theta_\usssB}
\newcommand{\thetaAB}{\theta_{\usssA\usssB}}

\newcommand{\Refc}[1]{Ref.~\cite{#1}}
\newcommand{\Refcs}[1]{Refs.~\cite{#1}}
\newcommand{\Eq}[1]{Eq.~\eqref{#1}}
\newcommand{\Eqs}[1]{Eqs.~\eqref{#1}}
\newcommand{\Fig}[1]{Fig.~\ref{#1}}
\newcommand{\Figs}[1]{Figs.~\ref{#1}}
\newcommand{\sectionc}[1]{Sec.~\ref{#1}}

\newcommand{\order}[1]{\mathcal{O}\!\left(#1\right)}

\newcommand{\dd}{\mathrm{d}} 

\newcommand{\GeV}{\mathrm{GeV}}

\newcommand{\Mpc}{\mathrm{Mpc}}

\newcommand{\OmegaCDM}{\Omega_\uc}
\newcommand{\OmegaB}{\Omega_\ub}

\newcommand{\thetaMC}{\theta_{\usssMC}}
\newcommand{\APSa}{A^{\usssPS}_{100}}
\newcommand{\APSb}{A^{\usssPS}_{143}}
\newcommand{\APSc}{A^{\usssPS}_{217}}
\newcommand{\rPSbc}{r^{\usssPS}_{143\times217}}
\newcommand{\ACIBb}{A^{\usssCIB}_{143}}
\newcommand{\ACIBc}{A^{\usssCIB}_{217}}
\newcommand{\rCIBbc}{r^{\usssCIB}_{143\times217}}
\newcommand{\gamCIB}{\gamma^{\usssCIB}}
\newcommand{\AtSZ}{A_{\utSZ}}
\newcommand{\AkSZ}{A_{\ukSZ}}
\newcommand{\xitSZCIB}{\xi^{\utSZ\times\usssCIB}}

\newcommand{\ca}{c_{100}}
\newcommand{\cc}{c_{217}}
\newcommand{\betaoo}{\beta_1^1}

\newcommand{\CltBB}{C_{\ell}^{\mathrm{BB,th}}}
\newcommand{\CloTT}{C_{\ell}^{\mathrm{TT,obs}}}
\newcommand{\CloTE}{C_{\ell}^{\mathrm{TE,obs}}}
\newcommand{\CloEE}{C_{\ell}^{\mathrm{EE,obs}}}
\newcommand{\CloBB}{C_{\ell}^{\mathrm{BB,obs}}}

\newcommand{\CloTB}{C_{\ell}^{\mathrm{TB,obs}}}

\newcommand{\tetan}{\theta_{\mathrm{n}}}
\newcommand{\tetalcdm}{\theta_{\mathrm{lcdm}}}
\newcommand{\tetalcdmMAX}{\theta_{\mathrm{lcdm}}^{\max}}
\newcommand{\tetas}{\theta_{\mathrm{s}}}
\newcommand{\tetai}{\theta_{\uinf}}
\newcommand{\tetar}{\theta_{\ureh}}

\newcommand{\Mp}{M_\usssPl}
\newcommand{\Mpl}{\Mp}
\newcommand{\vev}{\textit{vev}\xspace}

\newcommand{\CAMB}{\texttt{CAMB}\xspace}
\newcommand{\COSMOMC}{\texttt{COSMOMC}\xspace}
\newcommand{\ASPIC}{\texttt{ASPIC}\xspace}

\newcommand{\MULTINEST}{\texttt{MultiNest}\xspace}

\newcommand{\EI}{\textit{Encyclop{\ae}dia Inflationaris}\xspace}

\newcommand{\kstar}{k_*}
\newcommand{\etastar}{\eta_*}
\newcommand{\Pstar}{P_*}

\newcommand{\Hstar}{H}

\newcommand{\fnlloc}{f_{_\uNL}^\uloc}

\newcommand{\epsstar}[1]{\epsilon_{#1}}

\newcommand{\cS}{c_{_{\mathrm{S}}}}

\newcommand{\like}{\Like}
\newcommand{\likeinf}{\Like_\text{eff}}

\newcommand{\BayesFactor}[2]{B^{#1}_{#2}}
\newcommand{\Bref}[1]{\BayesFactor{#1}{\usssREF}}
\newcommand{\Bsr}[1]{\BayesFactor{#1}{\usssSR}}

\newcommand{\Cb}{\Complexity}

\newcommand{\Nevid}{193}

\newcommand{\Nmod}{N^{\umod}}

\newcommand{\Nuc}{N^{\uuc}}

\begin{document}

\title{Compatibility of Planck and BICEP2 results in light of inflation}

\author{J\'er\^ome Martin} \email{jmartin@iap.fr}
\affiliation{Institut d'Astrophysique de Paris, UMR 7095-CNRS,
Universit\'e Pierre et Marie Curie, 98 bis boulevard Arago, 75014
Paris, France}

\author{Christophe Ringeval} \email{christophe.ringeval@uclouvain.be}
\affiliation{Centre for Cosmology, Particle Physics and Phenomenology,
  Institute of Mathematics and Physics, Louvain University, 2 Chemin
  du Cyclotron, 1348 Louvain-la-Neuve (Belgium)}

\author{Roberto Trotta} \email{r.trotta@imperial.ac.uk}
\affiliation{Imperial College London, Astrophysics \& Imperial
  Centre for Inference and Cosmology, Blackett Laboratory, Prince
  Consort Road, London SW7 2AZ (United Kingdom)}

\author{Vincent Vennin} \email{vennin@iap.fr}
\affiliation{Institut d'Astrophysique de Paris, UMR 7095-CNRS,
Universit\'e Pierre et Marie Curie, 98 bis boulevard Arago, 75014
Paris, France}

\date{\today}

\begin{abstract}
   We investigate the implications for inflation of the detection of
   $B$-modes polarization in the Cosmic Microwave Background (CMB) by
   BICEP2. We show that the hypothesis of primordial origin of the
   measurement is favored only by the first four bandpowers, while the
   others would prefer unreasonably large values of the
   tensor-to-scalar ratio. Using only those four bandpowers, we carry
   out a complete analysis in the cosmological and inflationary
   slow-roll parameter space using the BICEP2 polarization
   measurements \emph{alone} and extract the Bayesian evidences and
   complexities for all the {\EI} models. This allows us to determine
   the most probable and simplest BICEP2 inflationary
   scenarios. Although this list contains the simplest monomial
   potentials, it also includes many other scenarios, suggesting that
   focusing model building efforts on large field models only is
   unjustified at this stage. We demonstrate that the sets of
   inflationary models preferred by Planck alone and BICEP2 alone are
   almost disjoint, indicating a clear tension between the two data
   sets.  We address this tension with a Bayesian measure of
   compatibility between BICEP2 and Planck. We find that for models
   favored by Planck the two data sets tend to be incompatible,
   whereas there is a moderate evidence of compatibility for the
   BICEP2 preferred models. As a result, it would be premature to draw
   any conclusion on the best Planck models, such as Starobinsky
   and/or K\"ahler moduli inflation. For the subset of scenarios not
   exhibiting data sets incompatibility, we update the evidences and
   complexities using both data sets together.
\end{abstract}


\maketitle

\section{Introduction}
\label{sec:intro}

The recent discovery of $B$-mode polarization in the Cosmic Microwave
Background (CMB) by BICEP2~\cite{Ade:2014xna}, if confirmed to be of
primordial origin~\cite{Liu:2014mpa}, would constitute a breakthrough
for our understanding of early universe cosmology. In addition to
lensing, $B$-mode can be generated by either vector perturbations or
tensor perturbations~\cite{Saga:2014zra}. Vectors do not propagate in
a Friedmann-Lema\^{\i}tre universe (see, however,
\Refc{Ringeval:2003na}) and can be a potential explanation of the
BICEP2 data only if they are incessantly generated by active sources
such as cosmic strings~\cite{Moss:2014cra, Lizarraga:2014eaa,
  Durrer:2014raa} or magnetic fields~\cite{Bonvin:2014xia}. These,
however, are severely constrained by other
measurements~\cite{Ringeval:2010ca, Ade:2013xla}.

\par

Tensor modes are a natural and expected outcome of cosmic inflation
although the uncertainty on their amplitude is huge (several orders of
magnitude). In this context, the BICEP2 result might represent the
first detection of primordial gravity waves produced in the early
Universe~\cite{Grishchuk:1974ny, Grishchuk:1975uf} and, therefore,
could give us precious information about the physical conditions that
prevailed at that time. Of course, the BICEP2 result needs to be
confirmed by other measurements before one can be sure that primordial
$B$-mode have really been detected. In this paper, our working
hypothesis will be that this is indeed the case. On general grounds,
it is anyway always interesting to explore the implications for
inflation of a non-negligible level of primordial gravity waves.

\par

The claimed amplitude of the signal corresponds to a tensor-to-scalar
ratio of $r=0.2^{+0.07}_{-0.05}$ or $r=0.16^{+0.06}_{-0.05}$ depending
on how polarized dust foregrounds are modeled and/or subtracted.
Recent works~\cite{Mortonson:2014bja} have, however, cast doubts on the
modeling of the foreground dust, which could potentially lead to the
amplitude of the tensor modes signal to be much lower, to the point of
becoming undetectable.  In the following, we shall take the BICEP2
result at face value, pending further investigation, most notably
thanks to the recently released Planck dust
maps~\cite{Ade:2014gna}. The BICEP2 measurement, if, as already
mentioned, interpreted as of primordial origin, has several important
physical consequences that we now discuss.

\begin{figure}
\begin{center}
\includegraphics[width=\wdblefig]{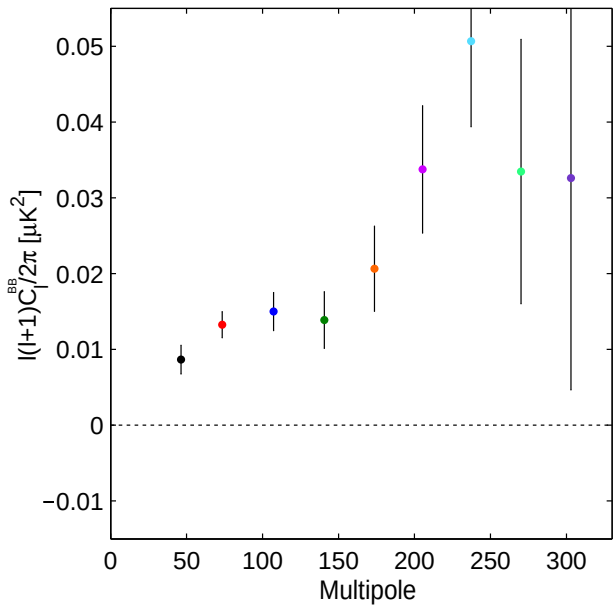}
\caption{$B$-mode angular power spectrum in the nine bandpowers
  measured by the BICEP2 experiment. Figure extracted from
  \Refc{Ade:2014xna}.}
\label{fig:bicep2BB}
\end{center}
\end{figure}

First, the energy scale of inflation~\cite{Starobinsky:1980te,
  Guth:1980zm, Linde:1981mu, Starobinsky:1982ee, Guth:1982ec,
  Albrecht:1982wi, Bardeen:1983qw, Linde:1984st, Mukhanov:1981xt, Ho:2014xza} is
fixed and roughly given by
\begin{equation}
\label{eq:energyinflation}
\rho^{1/4}\simeq 2.2 \left(\frac{r}{0.2}\right)^{1/4} 
10^{16} \, \GeV,
\end{equation}
\ie around the grand unified theory (GUT) energy scale. A more
accurate determination of this energy scale and the Hubble rate during
inflation are given in \sectionc{sec:energyscale}. Inflation is
therefore a high energy phenomenon by particle physics standard.

\par

Second, this result would favor that single field slow-roll
scenarios achieve the best compromise between quality of the fit and
theoretical simplicity~\cite{Martin:2013nzq}. Indeed, in more
complicated models, the tensor-to-scalar ratio is generically (but not
necessarily) smaller than in the standard case\footnote{With the
  notable exception of
  G-inflation~\cite{Kobayashi:2010cm,Kobayashi:2011pc} where $r=16\cS
  \sigma$, with $\sigma$ a complicated function of the inflaton field
  and its derivative, possibly larger than one even in the slow-roll
  limit.}. For instance, for
$K$-inflation~\cite{ArmendarizPicon:1999rj}, one has $r=-8\nT \cS$
where $\cS<1$ is the sound speed of the
fluctuations~\cite{Garriga:1999vw}. For two-field inflation, one can
write $r= -8\nT \sin ^2 \Theta\leq -8\nT$ , where $\sin \Theta$ is a
term taking into account the possible evolution of scalar modes on
super Hubble scales~\cite{Wands:2007bd}. For multiple field inflation,
the above equality becomes an inequality, namely $r\leq -8\nT \sin ^2
\Theta $, thus strengthening the argument presented before (up to the
special case of massive Nflation which inherits some of the properties
of a single $m^2 \phi^2$ model~\cite{Liddle:1998jc, Kim:2006ys,
  Piao:2006nm, Easther:2013rva}). Of course, this certainly does not
mean that these more complicated models are ruled out by BICEP2 (as a
matter of fact, they are not!), but together with the absence of
detection of isocurvature modes and primordial non-Gaussianities, this
reinforces the fact that they are not needed in order to give a
satisfactory description of the data. Clearly, this argument should be
toned down given that multiple field models are often well motivated
from a high energy point of view and, moreover, can predict a non-negligible $r$ even if the field excursion is smaller than the Planck
mass~\cite{Dimopoulos:2005ac} (see below). Also notice that, for the
simplest and preferred class of inflationary models mentioned above,
the non-Gaussianities are characterized by $\fnlloc=5(1-\nS)/12 \simeq
1.6\times 10^{-2}$~\cite{Maldacena:2002vr} since
Planck~\cite{Ade:2013zuv,Ade:2013uln} has measured
$\nS=0.9603\pm0.0073$. Therefore, unless one is able to reach the
$10^{-2}$ level, it seems impossible to measure $V(\phi)$ using the
precise shape of the three-point correlation function. The $10^{-2}$
level appears to be extremely challenging given our present day
capabilities and, as a consequence, this reinforces the importance of
a measurement of $r$ since this opens a realistic opportunity to
identify the correct inflationary scenario.

\par

Third, in the framework of single field slow-roll scenarios, the BICEP2
result implies a lower bound on the first Hubble flow function, which
is also given by $\epsilon_1\simeq \Mp^2(V_\phi/V)^2/2$. Therefore,
the first derivative of the inflaton potential can be
constrained. Furthermore, since the deviation from scale invariance
$\nS-1$ depends on a combination of the first and second derivatives
of the potential (at leading order in slow roll), this automatically
also provides a measurement of the second derivative of the
potential. It is also interesting to notice that a constraint on
$\epsilon_1$ does not modify our estimate of the importance of the
stochastic effects for CMB scales~\cite{Starobinsky:1986fx,
  Starobinsky:1994bd, Martin:2005ir}. Indeed, if $\Delta_{\uq}$ is the
typical quantum excursion of the inflaton field during one $e$-fold
and $\Delta_{\ucl}$ its classical excursion, then
$\Delta_\uq/\Delta_\ucl\simeq H/(\Mp \sqrt{\epsilon_1})\simeq
\sqrt{{\calP}_{\zeta}}\simeq 10^{-5}$. The point is that
$\Delta_\uq/\Delta_\ucl$ does not depend on $\epsilon_1$ alone but on
the combination $H/\sqrt{\epsilon_1}$ which was already measured
before BICEP2 since it turns out to be exactly the amplitude of the
scalar modes. However, a measurement of $r$ also gives indications
about the shape of the potential (see below) and, then,
$\Delta_\uq/\Delta_\ucl>1$ may become possible but necessarily outside
the observable window. If $r\simeq 0.2$ favors potentials for which
this systematically happens, one should still pay attention to how
measurements made on the CMB scales should be extrapolated to the part
of the inflaton's potential supporting a stochastic
regime~\cite{Kinney:2014jya}. Therefore, observationally speaking, the
question of knowing if non-perturbative quantum effects can play an
important role in the early Universe is still open~\cite{Linde:1986fc,
  Linde:1986fd, Goncharov:1987ir}.

\par

Fourth, the model building problem is also impacted by the BICEP2
result. Indeed, by definition of the first Hubble flow function, one
has $\Delta \phi/\Mp=\order{1}(r/0.2)^{1/2}$~\cite{Lyth:1996im,
  Antusch:2014cpa} which indicates that the excursion of the field
during inflation is necessarily super-Planckian. The single field
models discussed before are usually viewed as effective models only,
valid up to a cutoff $\Lambda$~\cite{Baumann:2014nda}. This scale
should be less than $\Mp$ since $\Mp$ is the cutoff of General
Relativity and larger than $H$ since the model should be able to
describe what happens during inflation. In the framework of effective
field theories, when physical effects beyond the cut-off are taken
into account, the total Lagrangian of a given inflationary model can
be expressed as $\calL=\dot{\phi}^2/2+V(\phi)+\sum_i
c_i\calO_i/\Lambda^{n_i-4}$, where $V(\phi)$ contains renormalizable
terms only and $\calO_i$ represents a higher order operator of
dimension $n_i> 4$ (possibly a non minimal kinetic term) the amplitude
of which is controlled by the coefficient $c_i$. When an inflationary
model is designed, it usually makes use of
$\calL=\dot{\phi}^2/2+V(\phi)$ only and the higher order operators are
neglected. The validity of this approximation is questionable because
of the following two problems. First, as mentioned above, a large
value of $r$ implies a large value of $\Delta\phi$ compared to the
Planck mass and the operators $\calO_i$ may no longer be
negligible. Solutions to these issues are either to fine-tune the
couplings between the light and heavy fields or to assume the
existence of a symmetry (typically the shift symmetry) to forbid the
dangerous higher order operators. But, then, this raises the question
of the origin of this symmetry in the full theory, that is to say the
question of the UV completion of the model.  For a nice and more
complete discussion on all these issues, see for instance
\Refc{Baumann:2014nda}.  Second, the parameters of $V(\phi)$ usually
get corrected by heavy field loops. For instance, a mass term
typically acquires the following form: $m^2\rightarrow m^2+gM^2\ln
(\Lambda /\mu)$, where $\mu $ is a renormalizable scale, $M>\Lambda $
the mass of a heavy field and $g$ the coupling between $\phi$ and the
heavy field\footnote{Notice that regularizing the loop integral with a
  cut-off would have produced a correction proportional to
  $\Lambda^2$, namely $m^2\rightarrow m^2+g\Lambda ^2$. However, this
  approach is not consistent as can be nicely illustrated on the
  example of the regularization of the cosmological constant. Indeed,
  if one regularizes $\rho_{\mathrm{vac}}$ with a cut-off, one obtains
  that $\rho _{\mathrm{vac}}\rightarrow
  \rho_{\mathrm{vac}}+\Lambda^4$. However, this method breaks Lorentz
  invariance and, as a consequence, one obtains the wrong equation of
  state, $w=p_{\mathrm{vac}}/\rho_{\mathrm{vac}}=1/3$ instead of
  $w=-1$. If, on the contrary, the loop integral is regularized with a
  method that respects Lorentz invariance (for instance dimensional
  regularization), then one obtains $\rho_{\mathrm{vac}}\rightarrow
  \rho_{\mathrm{vac}}+M^4\ln(\Lambda/\mu)$ and $w=-1$, see
  \Refc{Martin:2012bt}. In other words, if there is no new physics
  beyond the standard model, there is no hierarchy problem.}. This
means that the mass of the inflaton becomes larger than the Hubble
rate and that the potential is no longer flat enough to support
inflation. Notice however that this issue is, a priori, always present
even in a model where $r$ is small.

\par

Fifth, the BICEP2 result exacerbates the problems of inflationary
magnetogenesis~\cite{Turner:1987bw, Ratra:1991bn,
  Martin:2007ue}. Recent observations indicate the presence of
magnetic fields of strength ranging from $10^{-17}$ to $10^{-15}$
Gauss on megaparsec scales and such a large coherence length suggests
a cosmological origin~\cite{Essey:2010nd, Dolag:2010ni,
  Neronov:1900zz, Tavecchio:2010mk}. In order to produce a magnetic
field during inflation, one needs to break conformal invariance. For
instance, this can be achieved by considering a term $f^2(\phi)F_{\mu
  \nu} F^{\mu \nu}$. A simple parametrization for the function
$f(\eta)$ is given by $f\propto a^{\alpha}$~\cite{Martin:2007ue} since
the choices $\alpha\simeq 2$ or $\alpha \simeq -3$ both lead to a flat
spectrum. For $\alpha=2$ $f$ is a growing function of time, and since
the gauge field $A_\mu$ is also coupled to charged fermions with a
coupling constant $g_{\rm eff}(\eta)\propto g/f(\eta)$, this implies
that the system is in a non perturbative regime during
inflation~\cite{Demozzi:2009fu} (for a possible solution, see
\Refc{Ferreira:2013sqa, Ferreira:2014hma}). On the other hand, the
solution $\alpha=-3$ suffers from a backreaction problem. In order to
avoid a too important production of the electric field, the only way
out is then to lower the energy scale of inflation, \ie $H/\Mp
\lesssim 10^{-20}$~\cite{Martin:2007ue, Demozzi:2012wh,
  Ringeval:2013hfa}. The BICEP2 result would invalidate this solution
and, therefore, one is left in a situation where inflationary
magnetogenesis appears to be more problematic than before.

\par

Sixth, the detection of a quite large value of $r$ raises the
question of whether one can directly see the primordial gravitational
waves. With $r\simeq 0.2$ and $\nT \simeq -r/8 \simeq -0.025$, see
\Eq{eq:consistency}, one expects to have today $\Omega_{\ugw}\simeq
10^{-15}$ and experiments such as VIRGO~\cite{Accadia:2011ud} and
eLISA~\cite{AmaroSeoane:2012km} cannot detect such a tiny
signal. However, Japan's DECIGO~\cite{Kawamura:2011zz,Ando:2010zza},
Ultimate-DECIGO or NASA's Big Bang Observer
(BBO)~\cite{Crowder:2005nr} have a priori the sensitivity required to
directly probe the inflationary primordial gravity waves. Notice that
these experiments operate in the frequency range $f\in [10^{-2}\,
  \mbox{Hz},10\, \mbox{Hz}]$ and this could render the measurements of
the reheating parameters, such as the reheating temperature and/or the
equation of state parameter, feasible~\cite{Nakayama:2008wy,
  Kuroyanagi:2013ns}.

\par

Seventh, it has been claimed~\cite{Krauss:2013pha,
  Ashoorioon:2012kh,Markkanen:2014dba} that the BICEP2 results would
represent the first experimental evidence for quantum gravity since,
in the framework of inflation, the transverse and traceless component
of the perturbed metric is a quantum operator. This has indeed been
known for 40 years~\cite{Grishchuk:1974ny, Grishchuk:1975uf} and
more than 20 years in the context of
inflation~\cite{Mukhanov:1990me, Martin:2004um,
  Martin:2007bw}. However, this was already the case for scalar
modes~\cite{Mukhanov:1981xt, Mukhanov:1990me, Martin:2004um,
  Martin:2007bw}. Indeed, their equation of motion derives from the
perturbed quantum Einstein equations, $\delta \hat{G}_{\mu \nu}=8\pi
G\delta \hat{T}_{\mu \nu}$. To put it differently, the Mukhanov-Sasaki
quantum operator $\hat{v}$, that characterizes the amplitude of scalar
modes, is expressed in terms of the perturbed inflaton field $\delta
\hat{\phi}$ and the Bardeen potential $\hat{\Phi}$, concretely
$\hat{v}\equiv \delta \hat{\phi }+(\phi'/\calH)\hat{\psi}=\delta
\hat{\phi }^{({\rm gi})}+(\phi'/\calH)\hat{\Phi}$ (where $\delta
\hat{\phi}^{({\rm gi})}$ is the gauge invariant perturbed field and
$\hat{\psi}$ is the scalar component of the perturbed metric
proportional to $\delta_{ij}$). We see that the perturbed metric is
also a quantum operator in the scalar sector and is directly related
to the CMB anisotropies. Notice that a semi-classical formulation of
the problem, namely $\delta G_{\mu \nu}= 8\pi G\langle \delta
\hat{T}_{\mu \nu}\rangle $, does not help since $\delta \hat{T}_{\mu
  \nu}$, being by definition linear in $\delta \hat{\phi}$, satisfies
$\langle \delta \hat{T}_{\mu \nu}\rangle =0$. One might argue that the
scalar sector suffers from a gauge problem but this question has been
discussed and solved with the help of the gauge-invariant
formalism~\cite{Bardeen:1980kt}. There exists a gauge (the spatially
flat or uniform curvature gauge~\cite{Malik:2008im}) for which
$\psi=0$ and, therefore, $v=\delta \phi$. However, this cannot be used
as an argument that only field fluctuations must be quantized. Indeed,
there is another gauge (comoving orthogonal gauge~\cite{Malik:2008im})
where $\delta \phi=0$ and, hence, $v=(\phi'/\calH)\psi$. As a
consequence, the same logic leading to the above argument could also
be used to reach an opposite conclusion, namely that only metric
fluctuations (and not field perturbations) must be quantized. In fact,
as it is clear from the definition of the Mukhanov-Sasaki variable,
field and metric perturbations cannot be
disentangled~\cite{Riotto:2002yw} and the scalar modes are therefore
already a genuine signature of the quantum-mechanical nature of the
gravitational field. On the other hand, it is true that there still
exist open issues related to the quantum to classical transition of
these quantum fluctuations~\cite{Grishchuk:1990bj, Lesgourgues:1996jc,
  Burgess:2006jn, Kiefer:2008ku, Sudarsky:2009za, PintoNeto:2011ui,
  Martin:2012pea, Das:2013qwa, Das:2014ada}.

\par

Eighth, it is worth recalling that BICEP2 data do not only concern
the $B$-mode polarization but also the $E$-modes ($\CloTT$ and
$\CloTE$ are not yet publicly available). The fact that the
polarization spectrum $\CloEE$ has also been measured is fortunate
since it allows us to constrain scalar perturbations, and cosmology,
with the BICEP2 data alone~\cite{Galli:2014kla}. This is discussed
further in the following. Although not public, the BICEP2 team reports
a $\CloTB$ consistent with zero and this is relevant for models
containing a gravitational Chern-Simons
term~\cite{Lue:1998mq,Alexander:2004wk,Contaldi:2008yz}. This term is
necessarily present since it is generated by quantum corrections and
is generic in string theory. This implies that the two polarization
states of a gravitational wave behave differently. As a consequence,
the tensor-to-scalar ratio is modified and can even be
enhanced~\cite{Alexander:2004wk} (to be fair, a calculation of $r$ in
a regime where the enhancement is large remains very challenging).

\par

Another question which arises after BICEP2 is the implications of
these new data with regards to the shape of the inflaton potential
$V(\phi)$ and whether these implications are compatible with the
conclusions reached previously, and notably from Planck
data~\cite{Martin:2014vha, Martin:2013nzq,
  Martin:2013gra,Dorn:2014kga}. Let us recall that, given Planck
data, the best models in terms of evidences and complexities are such
that their potential is of the plateau type, the prototypical example
being the Starobinsky model~\cite{Starobinsky:1980te}. In more
quantitative terms, if one uses the Jeffreys'
scale~\cite{Trotta:2005ar, Trotta:2008qt} and counts the number of
models in the ``inconclusive'', ``weak evidence'', ``moderate
evidence'' and ``strong evidence'' zones with respect to the best, one
finds $26\%$ in the first category (corresponding to $17$ different
shapes of the potential), $21\%$ in the second, $17\%$ in the third
and, finally, $34\%$ in the fourth and last one. These numbers can be
further improved by another statistical indicator. If we restrict
ourselves to models having a number of unconstrained parameters
between zero and one, then only $9\%$ of the scenarios are preferred,
corresponding this time to $9$ different potentials. And these $9$
potentials are all of the plateau type. On the other hand, the
Jeffreys scale has to be taken as indicative, and it is usually considered
that only the models belonging to the strong evidence category (here,
$34\%$) can really be considered as robustly ``ruled out''. Therefore, we see
that the Planck data have been able to narrow down our theoretical
uncertainties efficiently and to point to a particular type of
potentials. As a consequence, an important question is whether the
BICEP2 measurements are in agreement with these conclusions and, more
generally, whether BICEP2 is compatible with Planck in the framework of single field slow-roll inflation.

\par

The present article aims at discussing the issues presented above. As
many inflationary models genuinely produce a small amount of tensor
modes, one would expect the BICEP2 data to severely cut a large volume
of the model space, thereby improving our knowledge of inflation
compared to what has already been established with Planck
data. However, one has first to address and quantify the compatibility
between BICEP2 and Planck data. For this, it is required to first
investigate both data sets independently. This may seem problematic
for BICEP2 as $B$-modes alone do not give constraints on the scalar
perturbations. But, as we show below, using both the $E$- and $B$
modes polarization measurements in only four bandpowers already gives
non-trivial constraints on both the standard cosmological parameters
and the primordial ones (see also \Refc{Galli:2014kla}). This allows
us to derive the evidences and complexities of all the {\EI} models
using BICEP2 data alone and thoroughly discuss the compatibility of
Planck and BICEP2 using the so-called
$\calR$-factors~\cite{Hobson:2002zf, Marshall:2004zd, Feroz:2008wr,
  Cabrera:2010xx, Feroz:2011bj, Arina:2012dr, Arina:2013jma}. These
are the Bayes factors giving the ratio between the probability of
compatibility to the probability of incompatibility assuming a given
model. By evaluating $\calR$ for slow-roll inflation and for each
{\EI} scenarios, one can determine the subset of models for which
Planck and BICEP2 data can be meaningfully combined to obtain
evidences and complexities from the joint data sets.

\par

This article is organized as follows. In the next section, we briefly
describe the method used to compute the Bayesian evidence of any
slow-roll inflationary model. In particular, our method is based on
the determination of an effective likelihood for inflation which
requires a slow-roll analysis of the Planck and BICEP2 data. The
results of the analysis for Planck can be found in
\Refc{Martin:2013nzq} and we present in \sectionc{sec:data} new
results for BICEP2 alone, and BICEP2 combined with Planck. In
particular, we discuss the compatibility of the data sets under the
hypothesis of slow-roll. In \sectionc{sec:result}, we present the
evidences and complexities for all the {\EI} models stemming from the
BICEP2 data alone and discuss what are the inflaton potential shapes
favored and how they differ from the Planck results. We then move on
to the compatibility of Planck and BICEP2 model by model and present
joint evidences and complexities for the scenarios under which both
data sets are not incompatible. Finally, in Sec.~\ref{sec:conclusion},
we summarize our findings and present our conclusions.

\section{Methodology}
\label{sec:method}

\subsection{Bayesian Evidence and Complexity}
\label{sec:evidcomp}

In this section, we briefly present the statistical methodology
adopted here to compute the Bayesian evidence and complexity for each
of the \EI models that, in the following, we denote by $\calM_i$.

The Bayesian evidence, given data $D$, is defined
by~\cite{Trotta:2008qt}
\begin{equation}
P(D \vert \calM_i) \equiv \calE \left(D\vert \calM_i\right)=
\int {\rm d}\theta_{ij}\like \left(\theta_{ij}\right)
\pi\left(\theta_{ij}\vert \calM_i\right),
\label{eq:defevidence}
\end{equation}
where $\theta_{ij}$ represents the parameters characterizing the model
$\calM_i$, $\like$ is the likelihood function (to be discussed below)
and $\pi\left(\theta_{ij}\vert \calM_i\right)$ is the prior
distribution for the parameter $\theta_{ij}$. As usual in Bayesian
analysis, the choice of the priors plays a crucial role and a complete
study of the $\pi\left(\theta_{ij}\vert \calM_i\right)$ for all the
\EI models can be found in \Refc{Martin:2013nzq}. Here, we adopt
the same choices. The Bayesian complexity can be expressed
as~\cite{Kunz:2006mc}
\begin{equation}
\label{eq:defcomplex}
\Cb_i=\left \langle -2\log \like \left(\theta_{ij}\right)\right \rangle
+2\log \like \left(\theta_{ij}^{\rm ML}\right),
\end{equation}
where $\langle \cdot \rangle$ means averaging over the posteriors and
$\theta_{ij}^{\rm ML}$ represents the maximum likelihood estimate of
the model's parameters. 

The Bayesian evidences are often normalized to a reference model
$\calM_\usssREF$ and one defines $\Bref{i}\equiv \calE(D\vert\calM_i)
/\calE(D\vert \calM_\usssREF)$. They give us information about the posterior
probability of the model $\calM_i$ (for non-committal model priors),
\begin{equation}
P\left(\calM_i\vert D\right)=\frac{\Bref{i}}{\sum _j\Bref{j}}\,.
\end{equation}
On the other hand, the Bayesian complexities tell us something about the number of
unconstrained parameters
\begin{equation}
\label{eq:defuparam}
\Nuc_i \equiv N_i-\Cb_i,
\end{equation}
where $N_i$ is the total number of parameters of the model under
scrutiny. The above considerations show that, given a data set
$D$, the performance of a model can be described by the numbers
$(\Nuc,\Bref{})$.

\subsection{Compatibility of data sets}
\label{sec:combining}

Although the previous discussion is readily applicable for either the
Planck ($\Dp$) or BICEP2 data ($\Db$) separately, computing a joint
evidence from BICEP2 and Planck, namely using $D = \{\Dp, \Db\}$,
requires some precaution. Indeed, it is crucial to determine whether a
small value of $P(D \vert \calM_i)$ is the consequence of $\calM_i$
being a poor explanation of the data, or whether this results from the
tension between Planck and BICEP2.

As detailed below, there is a some tension between the two data sets,
when interpreted in terms of tensor modes amplitude. Combining the two
data sets blindly could potentially lead to a joint likelihood
function that peaks in a region of parameter space that is not
favored by either experiment--an obviously undesirable situation.

In order to study the compatibility of BICEP2 and Planck, we resort to
a Bayesian measure defined as follows~\cite{Hobson:2002zf,
  Marshall:2004zd, Feroz:2008wr, Cabrera:2010xx, Feroz:2011bj,
  Arina:2012dr, Arina:2013jya, Arina:2013jma}:
\begin{equation}
\Rfac_i \equiv \frac{P(\Dp, \Db | \calM_i, \calH_\text{c})}{P(\Dp | \calM_i, \calH_\text{ic}) P(\Db |
  \calM_i, \calH_\text{ic})}\, .
\label{eq:Rfacdef}
\end{equation}
This quantity represents the posterior between the hypothesis that the
two data sets are compatible, ($\calH_\text{c}$, when $\Rfac_i > 1$) versus the hypothesis
that they are not ($\calH_\text{ic}$ and thus described by different sets of parameters, when $\Rfac_i < 1$), assuming the inflationary model
$\calM_i$ and noncommittal priors between the two hypotheses, $P(\calH_\text{c}) = P(\calH_\text{ic}) = 1/2$. Various proto-typical situations illustrating the behavior
or $\calR$ are presented in the appendix (see \sectionc{sec:toy}),
where one can gain some insight on why $\Rfac$ measures
compatibility. The $\calR$ factor can also be reexpressed in terms of
the conditional predictive probability for BICEP2 data, by noting that
\begin{equation}
P(\Dp, \Db | \calM_i) = P(\Db | \Dp, \calM_i) P(\Dp | \calM_i).
\label{eq:jointP}
\end{equation}
Using Eq.~\eqref{eq:jointP}, we obtain the simpler expression
\begin{equation}
\Rfac_i = \frac{P(\Db | \Dp, \calM_i)}{P(\Db | \calM_i)}\,,
\end{equation}
which shows that $\Rfac_i$ is large if the probability of obtaining
data $\Db$, given the Planck data $\Dp$, is large.

\subsection{Likelihood Functions}
\label{sec:likelihood}

\begin{figure}
\begin{center}
\includegraphics[width=\wdblefig]{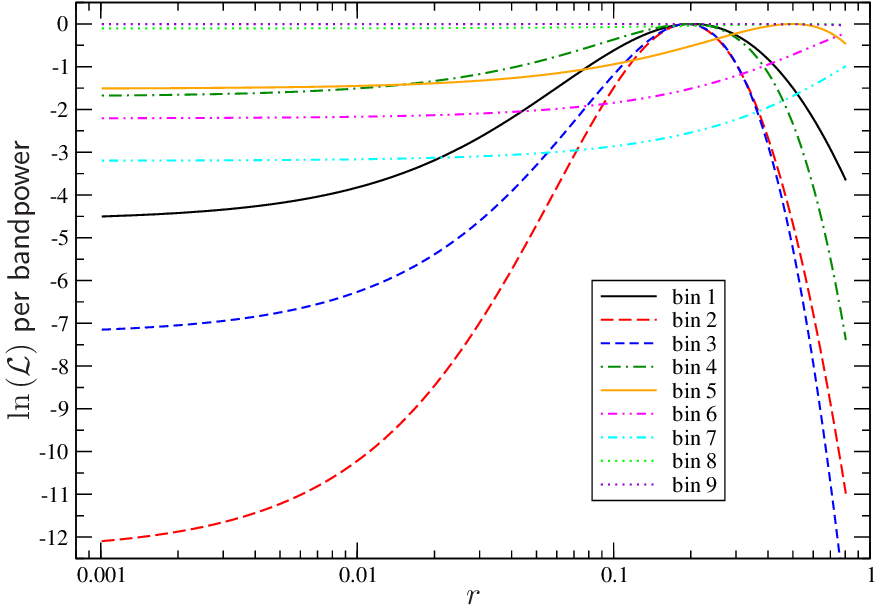}
\end{center}
\caption{BICEP2 likelihood function over the tensor-to-scalar ratio
  $r$, assuming power law primordial spectra and with the $\Lambda$CDM
  parameters fixed to their mean value obtained from the Planck data
  alone. $B$-modes from lensing are included and we have presented the
  contribution of each of the nine BICEP2 bandpowers. Only the first
  four bandpowers favor the hypothesis of primordial tensor modes of
  inflationary origin whereas the others exhibit a likelihood maximal
  at problematic large values of $r$.}
\label{fig:likebins}
\end{figure}

The likelihoods considered in the following have been provided by the
Planck Collaboration~\cite{Ade:2013kta} and the BICEP2
team~\cite{Hamimeche:2008ai, Chiang:2009xsa, Ade:2014xna}. Concerning
the Planck likelihood, we have used the ``CamSpec'' likelihood for the
temperature power spectrum in the multipole range $50<\ell<2500$
complemented with the ``Commander'' likelihood for
$2<\ell<49$. Moreover, following the data analysis method of
\Refcs{Ade:2013ktc, Ade:2013zuv}, we have also used the WMAP
polarization data for $\ell \le 32$~\cite{Dunkley:2008ie,
  Hinshaw:2012aka, Bennett:2012zja}. These data sets are the same as
the ones used in \Refc{Martin:2013nzq}. Concerning the BICEP2
likelihood, we have written a FORTRAN code from scratch based on the
approximation of \Refc{Hamimeche:2008ai} and as implemented by the
BICEP2 team (see \Refc{Barkats:2013jfa}). Our results are identical to
the ones obtained with the latest version of
{\COSMOMC}~\cite{Lewis:2002ah} in which the BICEP2 likelihood has also
been implemented. The BICEP2 measurements are publicly
available\footnote{See \url{http://bicepkeck.org}.}.

As discussed in \Refc{Ade:2014xna}, when assuming primordial power law
power spectra, the BICEP2 likelihood for the tensor-to-scalar ratio
$r$ peaks at a value around $0.2$, which is significantly larger than
those favored by the Planck data. In \Fig{fig:likebins}, we have
represented the BICEP2 likelihood profile along $r$, in each
bandpower, when the $\Lambda$CDM cosmological parameters are fixed to
their mean values obtained from the Planck data
alone~\cite{Ade:2013zuv}. Let us notice that the likelihood has been
estimated using {\CAMB}~\cite{Lewis:1999bs} to provide the expected
$\CltBB$ for each value of $r$ while including the lensing effects
which convert $E$-modes into $B$-modes. This figure shows that over
the nine bandpowers provided by the BICEP2 team, the second bin
carries most of the statistical weight and, moreover, only four bins
are reasonable with the hypothesis that the measured $\CloBB$ are
sourced by tensor modes of inflationary origin. Indeed, already for
the bandpower five, the likelihood peaks at a value $r>0.5$. The
bandpowers six to nine would even favor a tensor-to-scalar ratio
larger than one. Those bandpowers do not significantly weigh in the
total likelihood as their associated errors are relatively large (see
\Fig{fig:bicep2BB}). However, as they seem to suffer from a systematic
excess, the origin of which not being inflation or lensing, we
have decided to perform our data analysis using only the first four
bandpowers of the BICEP2 data hoping that they are not too much
affected by such systematics. In fact, we have also checked that
including all the bins in the analysis does not modify in a
substantial way our conclusions.

\subsection{Fast Evidence Computation}
\label{eq:fast}

Given our likelihood function, we briefly summarize in this section
how the Bayesian evidence of a given {\EI} model can be fast
computed.

\par

Any inflationary model is characterized by the parameters $\tetai$
describing the shape of the potential [for instance, for large field
  inflation where $V(\phi)=M^4(\phi/\Mp)^p$, one has $\tetai=(M,p)$]
and by the priors choice on those parameters~\cite{Martin:2013nzq}. We
also need parameters describing the reheating phase, $\tetar$, such as
the reheating temperature and the equation of state. In fact, one can
show that only one parameter is sufficient, the so-called rescaled
reheating parameter $R$~\cite{Turner:1983he, Martin:2006rs,
  Ringeval:2007am, Martin:2010kz, Easther:2011yq}. In the present
paper, following \Refc{Martin:2013nzq}, a Jeffreys' prior is assumed
such that $\tetar = \ln R\in [-46,15]$. Finally, the parameters
describing the post-inflationary phase are the standard cosmological
parameters associated with a $\Lambda$CDM model, plus the
astrophysical parameters entering the likelihood function. Those are
referred to as $\tetas$ in the following. As a consequence, the
evidence in \Eq{eq:defevidence}, for a model $\calM_i$, becomes
\begin{equation}
\begin{aligned}
\label{eq:evidence}
\calE\left(D\vert \calM_i\right) =\int  & \dd \tetas \dd \tetar 
\dd \tetai \calL\left(\tetas,\tetar, \tetai \right) 
\\ & \times \pi(\tetas)\pi(\tetar)\pi(\tetai),
\end{aligned}
\end{equation}
where $\pi$ represent the priors. The key remark here is to notice
that, as opposed to the cosmological and astrophysics parameters,
$\tetar$ and $\tetai$ affect the likelihood by modifying \emph{only}
the scalar $\calP_\zeta(k)$ and tensor $\calP_h(k)$ primordial power
spectra. As a consequence, one possibility would be to numerically
evaluate, for each inflationary model, $\calP_\zeta$ and
$\calP_h$~\cite{Martin:2010hh, Easther:2011yq, Ade:2013uln}. This is,
however, very time consuming.

Here, we rather choose to use the method developed in
\Refc{Ringeval:2013lea}. The main idea of this article is to bypass
any mode integration by modeling through a small number of parameters
the shape of the primordial spectra. Since we are only focused on
slow-roll inflation, we consider the second order slow-roll expansion
of the scalar and tensor primordial spectra around a pivot scale
$\kstar$~\cite{Stewart:1993bc, Gong:2001he, Martin:2002vn,
  Habib:2002yi, Schwarz:2001vv, Leach:2002ar, Casadio:2004ru,
  Casadio:2005xv, Casadio:2005em, Lorenz:2008et, Martin:2013uma,
  Jimenez:2013xwa}, namely
\begin{equation}
\begin{aligned}
  \calP_\zeta &= \dfrac{\Hstar^2}{8 \pi^2 \Mpl^2 \epsstar{1}} \bigg\{ 1
-2(1+C)\epsstar{1} - C \epsstar{2} \\ & + \left(\dfrac{\pi^2}{2} -3 +
2C + 2C^2 \right) \epsstar{1}^2 +\left(\dfrac{\pi^2}{24} -
\dfrac{C^2}{2} \right) \epsstar{2} \epsstar{3} \\ & +  \left( \dfrac{7
  \pi^2}{12} -6 -C+C^2\right) \epsstar{1} \epsstar{2} +
\left(\dfrac{\pi^2}{8} -1 +\dfrac{C^2}{2} \right) \epsstar{2}^2 \\ & +
\bigg[-2 \epsstar{1} - \epsstar{2} +(2+4C) \epsstar{1}^2 + (-1 + 2C)
  \epsstar{1} \epsstar{2} \\&+ C \epsstar{2}^2 - C \epsstar{2}
  \epsstar{3} \bigg] \ln\left(\dfrac{k}{\kstar}\right) + \bigg(2
  \epsstar{1}^2 + \epsstar{1} \epsstar{2} + \dfrac{1}{2} \epsstar{2}^2
  \\ & - \dfrac{1}{2} \epsstar{2} \epsstar{3} \bigg) \ln^2
\left(\dfrac{k}{\kstar}\right) \bigg\}
\label{eq:pzeta}
\end{aligned}
\end{equation}
and
\begin{equation}
\begin{aligned}
\calP_h &= \dfrac{2 \Hstar^2}{\pi^2 \Mpl^2} \\ & \times \left\{ 1 -
  2(1+C)\epsstar{1} +\left(\dfrac{\pi^2}{2}-3 + 2C +2C^2 \right)
  \epsstar{1}^2 \right.\\ & + \left. \left(\dfrac{\pi^2}{12} -2 -2C
  -C^2 \right) \epsstar{1} \epsstar{2} + \left[ -2 \epsstar{1} +
    (2+4C) \epsstar{1}^2 \right. \right. \\ &- \left. \left. 2(1+C)
    \epsstar{1} \epsstar{2} \right] \ln\left(\dfrac{k}{\kstar}\right)
  + \left(2 \epsstar{1}^2 - \epsstar{1}\epsstar{2} \right)
  \ln^2\left(\dfrac{k}{\kstar}\right) \right\},
\end{aligned}
\label{eq:ph}
\end{equation}
where $C=\gamma+\ln 2-2\simeq -0.72$, $\gamma $ being the Euler
constant. The quantities $\epsilon_n$ are the Hubble-flow parameters
evaluated at pivot Hubble exit, {\ie} at the conformal time $\etastar$
solution of $\kstar \etastar = -1$. The Hubble parameter $\Hstar$
entering the normalization is also evaluated at $\etastar$. Let us
notice that, by definition, $\Pstar \equiv \calP_\zeta(\kstar)$ is a
well-measured quantity which fixes the amplitude of the CMB
anisotropies. Let us also remark that, a priori, $\Pstar$ is not
directly proportional to $\CloTT$ since, when the tensor-to-scalar
ratio does not vanish, part of the signal also comes from
$\calP_h(\kstar)$. However, for our choice of pivot scale, $\kstar =
0.05\,\Mpc^{-1}$, the gravity waves contribution is already very
small.

The inflationary model dependence now only appears through the
explicit functionals $\epsilon_n(\tetar,\tetai)$. These are explicitly
derived for all models of the {\EI} in \Refc{Martin:2014vha} and can
be computed using the public library {\ASPIC}\footnote{\url{http://theory.physics.unige.ch/~ringeval/aspic.html}}. In other words, the
power spectra obtained in this way differ for different models because
the functionals $\epsilon_n(\tetar,\tetai)$ depend on the inflationary
model considered. Then, the Bayesian evidence can be obtained from
\Eq{eq:evidence} by marginalizing over all parameters, {\ie}
\begin{equation}
\begin{aligned}
\calE \left(D\vert \calM_i\right)&=\int  
\likeinf\left[\Pstar(\tetar, \tetai),\epsilon_n(\tetar, \tetai)\right]
\\ & \times \pi(\tetar)\pi(\tetai) \, \dd \tetar \dd \tetai,
\end{aligned}
\label{eq:evidence2}
\end{equation}
where we have defined the effective likelihood by
\begin{equation}
\begin{aligned}
\likeinf\left(\Pstar,\epsstar{n}\right) \equiv \int
\like\left(\tetas,\Pstar,\epsstar{n} \right) \pi(\tetas) \,\dd \tetas.
\end{aligned}
\label{eq:likeinf}
\end{equation}
The effective likelihood for inflation $\likeinf$ is the full
likelihood $\like$ marginalized over all the cosmological and
astrophysics parameters. Its estimation therefore requires a complete
data analysis that we present in the following. However, this has to
be done once and for all as the evidences of all the inflationary
models can be computed afterwards from \Eq{eq:evidence2}. In
practice, the functional shape of $\likeinf(\Pstar,\epsstar{n})$ is
fitted using a neural network interpolator allowing its very fast
evaluation.

\section{Data analysis}
\label{sec:data}

\begin{figure}
\begin{center}
\includegraphics[width=\wdblefig]{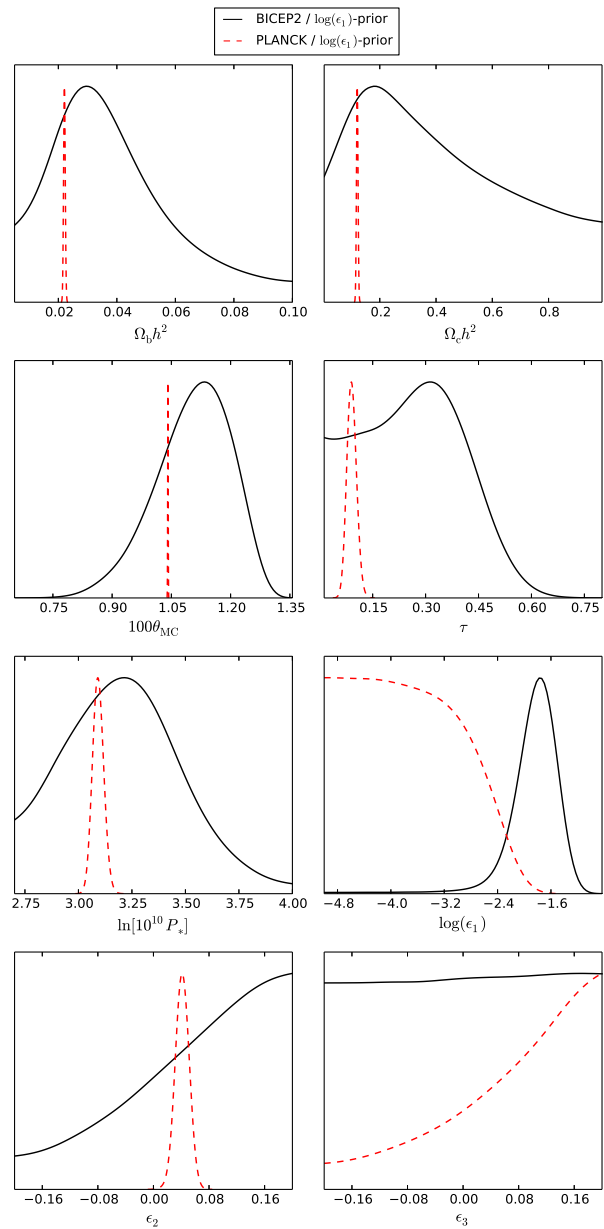}
\caption{One-dimensional marginalized posterior probability
  distributions for the cosmological and primordial slow-roll
  parameters obtained with BICEP2 data alone (solid black lines)
  compared to the corresponding Planck's posteriors (dashed red lines).}
\label{fig:b2sr}
\end{center}
\end{figure}

\begin{figure}
\begin{center}
\includegraphics[width=\wdblefig]{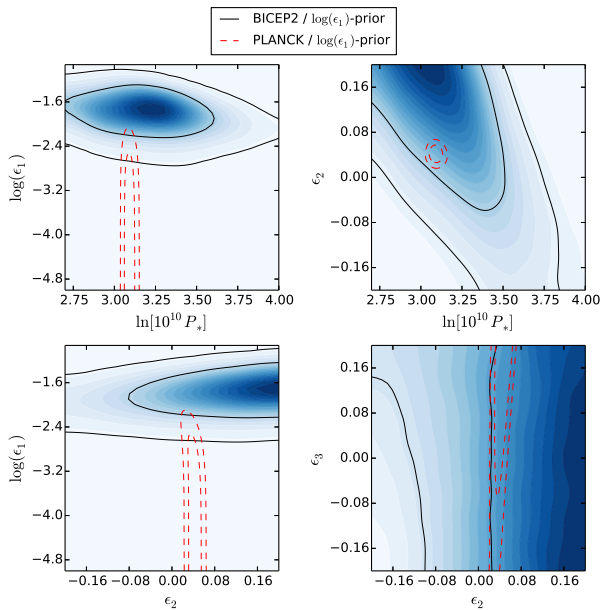}
\caption{Two-dimensional marginalized posterior probability
  distributions for the primordial slow-roll parameters obtained from
  BICEP2 data alone (solid black lines) compared to the corresponding
  Planck's posteriors (dashed red lines). The blue shading
  density traces the mean likelihood values for BICEP2
  (Jeffreys' prior on $\epsstar{1}$).}
\label{fig:b2sr2D}
\end{center}
\end{figure}

\begin{figure*}
\begin{center}
\includegraphics[width=\wsingfig]{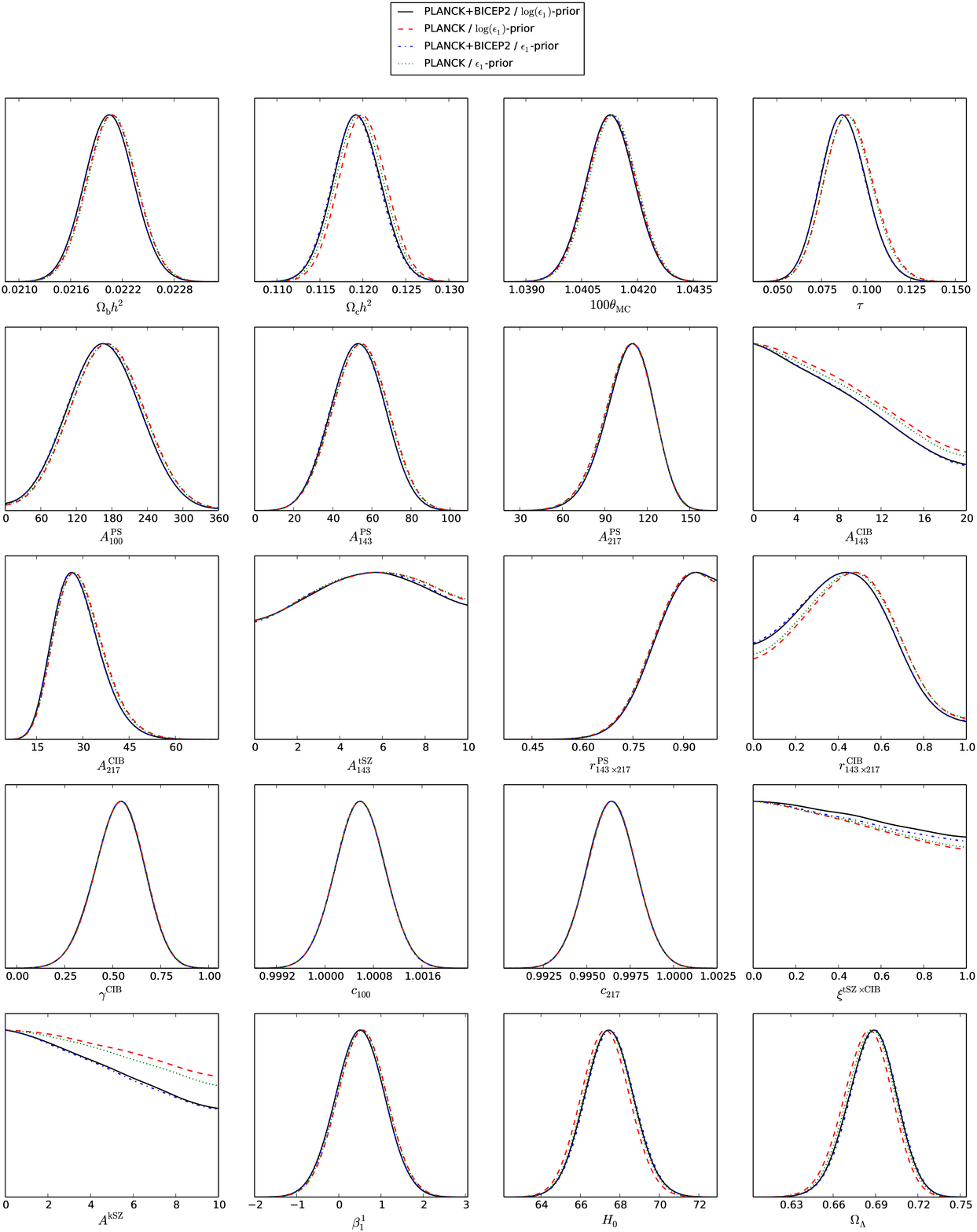}
\caption{One-dimensional marginalized posterior probability
  distributions on cosmological and astrophysics parameters associated
  with primordial power spectra having a second order slow-roll
  functional shape. These posteriors are robust against the four cases
  represented: Planck data alone, Planck and BICEP2 data combined,
  Jeffreys' prior on $\epsstar{1}$ or flat prior on $\epsstar{1}$.}
\label{fig:tetas}
\end{center}
\end{figure*}

\begin{figure}
\begin{center}
\includegraphics[width=\wdblefig]{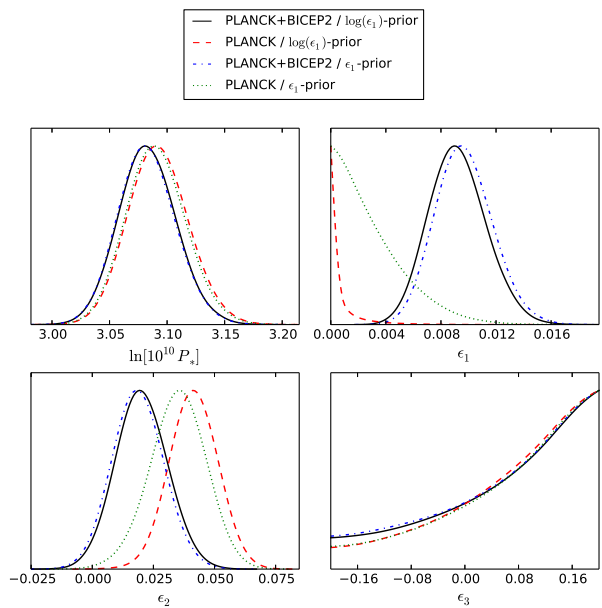}
\caption{One-dimensional marginalized posterior probability
  distributions for the primordial slow-roll parameters obtained with
  Planck and Planck plus BICEP2 data; with a Jeffreys' prior on
  $\epsstar{1}$ or a flat prior on $\epsstar{1}$. Notice the one-sigma
  shift of the $\epsstar{2}$ posterior towards smaller values when the
  BICEP2 data are included.}
\label{fig:bcpsr}
\end{center}
\end{figure}

\begin{figure}
\begin{center}
\includegraphics[width=\wdblefig]{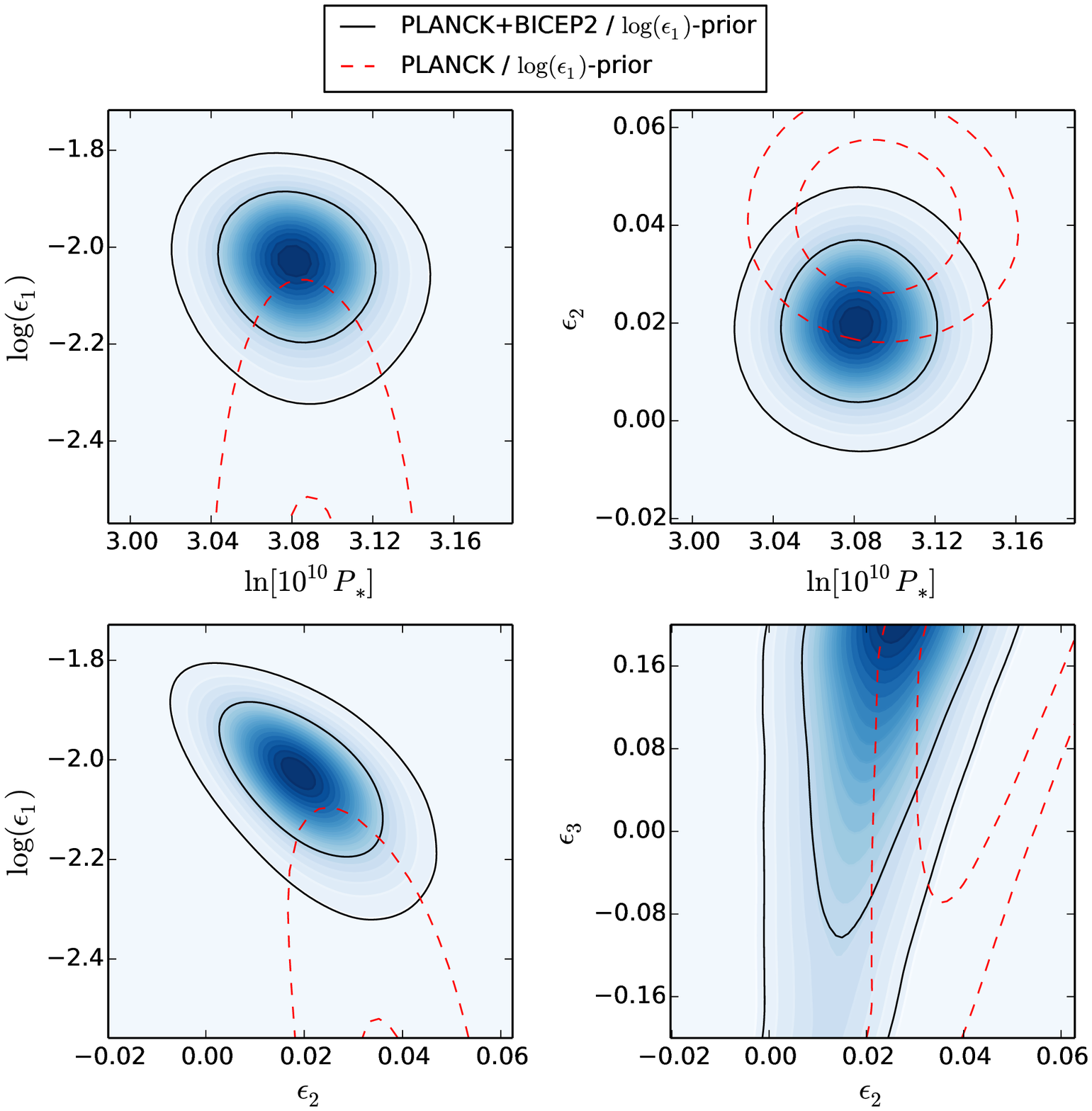}
\includegraphics[width=\wdblefig]{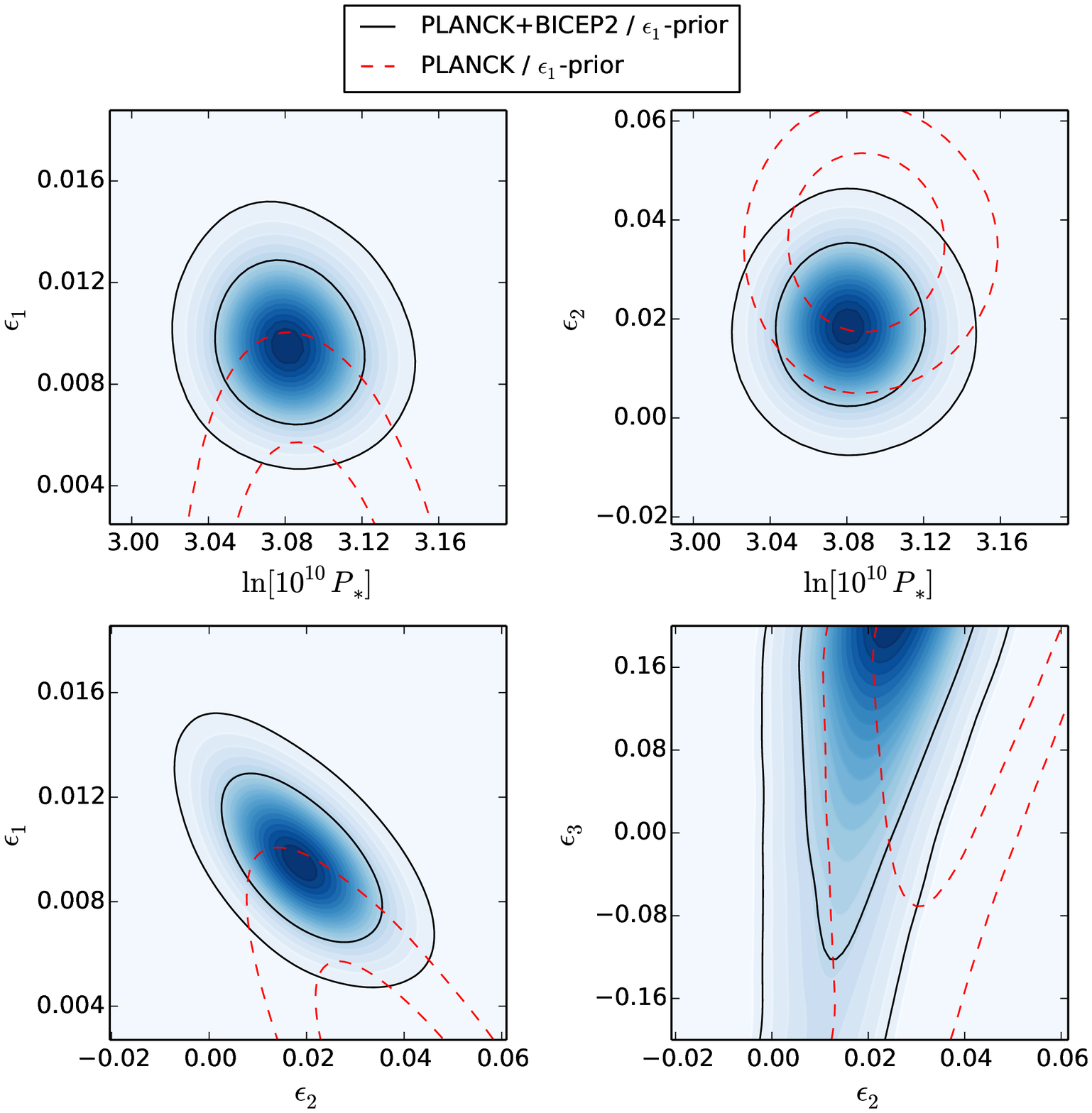}
\caption{One and two-sigma contour of the marginalized
  posterior probability distributions for the primordial slow-roll
  parameters obtained with Planck and Planck+BICEP2 data. The blue shading
  density traces the mean likelihood values for Planck+BICEP2
  (Jeffreys' prior on $\epsstar{1}$). The tension between Planck and
  Planck+BICEP2 data induces a $1.5$-sigma shift of the
  $\log(\epsstar{1})$ posterior towards higher values while shifting
  by one sigma the posterior of $\epsstar{2}$ towards zero.}
\label{fig:bcpsr_2D}
\end{center}
\end{figure}

In order to determine $\likeinf$, we have performed a
Markov-Chain-Monte-Carlo (MCMC) exploration of the slow-roll parameter
space using the BICEP2 and Planck likelihoods described above.

\subsection{Constraints from BICEP2}
\label{sec:srpostb2}

In this first section, we derive constraints on the cosmological
parameters using the BICEP2 data alone. The post-inflationary
universe, assumed to be a flat $\Lambda$CDM model, is described by the
parameters $\tetas$:
\begin{equation}
\begin{aligned}
\tetas & = \left(\OmegaB h^2, \OmegaCDM h^2, \tau,
100\thetaMC \right).
\label{eq:camparamsb2}
\end{aligned}
\end{equation}
The cosmological parameters are the baryons energy density (normalized
to the critical energy density) $\OmegaB$, the cold dark matter energy
density $\OmegaCDM$, the reduced Hubble parameter today $h$, the
optical depth $\tau$ to last scattering and an angle, $\thetaMC$,
related to the angular size of the sound horizon on the last
scattering surface~\cite{Lewis:2002ah}. The MCMC analysis was done by
means of the public code $\COSMOMC$~\cite{Lewis:2002ah} and a modified
version of the $\CAMB$ code~\cite{Lewis:1999bs} taking into account
that the initial power spectra are not simple power laws but are given
by the expressions~(\ref{eq:pzeta}) and~(\ref{eq:ph}). The priors on
the standard parameters are chosen in accordance with
\Refc{Ade:2013kta}. For the primordial parameters, we take a Jeffreys'
prior for $\ln \left(10^{10} \Pstar\right) \in[2.7,4.0]$ and for
$\epsstar{1}$, namely $\log(\epsstar{1})\in [-5,-0.7]$. For the other
slow-roll parameters, we choose flat priors on $\epsstar{2}$ and
$\epsstar{3}$ in $[-0.2,0.2]$. As already mentioned, the pivot scale
is chosen at $\kstar = 0.05\,\Mpc^{-1}$. These priors are the most
uninformative within slow-roll inflation. Indeed, the order of
magnitude of $\epsilon_1$ (which is always positive) is a priori
unknown as many models produce a level of tensor modes that can be
extremely small. This prior was the one assumed in
\Refc{Martin:2013nzq}

In Fig.~\ref{fig:b2sr}, we have represented the one-dimensional
marginalized posterior probability distributions for the standard and
slow-roll parameters obtained with BICEP2 data alone (solid black
lines) compared with the distributions inferred from Planck (dashed
red lines). It does not come as a surprise to see that, as long as the
$\tetas$'s are concerned, BICEP2 is much less constraining than
Planck. Concerning the primordial parameters, we see that BICEP2
measures $\epsilon_1$ (or $r$) since it is sensitive to both the
amplitude of the tensor power spectrum through $\CloBB$ and the
amplitude of the scalar power spectrum through $\CloEE$. The quantity
$P_*$, that is to say $\calP(\kstar)$, is indeed constrained as can be
seen on the figure. On the other hand, the second and third slow-roll
parameters $\epsilon_2$ and $\epsilon_3$ are not constrained at all.

\par

Fig.~\ref{fig:b2sr2D} shows the two-dimensional posterior probability
distributions in the primordial parameter space from BICEP2 alone
(solid black contours) and from Planck alone (red dashed
contours). The upper and lower left panels are especially interesting
since they illustrate the existing tension between the BICEP2
and Planck data in the sense that the one-sigma contours do not
overlap (while the two-sigma contours do). Unsurprisingly, the first
slow-roll parameter is well determined by BICEP2 while the second is
well constrained by Planck.

\subsection{Constraints from BICEP2 and Planck}
\label{sec:srpost}

We now turn to the joint analysis where the BICEP2 and Planck data are
simultaneously considered. The post-inflationary universe is, as
before, a flat $\Lambda$CDM model, and is now described by a
larger set of parameters $\tetas$:
\begin{equation}
\begin{aligned}
\tetas & = \left(\OmegaB h^2, \OmegaCDM h^2, \tau,
100\thetaMC, \APSa, \APSb, \APSc, \rPSbc, \right. \\ & \left. \ACIBb,
\ACIBc, \rCIBbc, \gamCIB,\AtSZ, \AkSZ, \xitSZCIB,
\ca, \right. \\ & \left.
\cc, \betaoo \right).
\label{eq:camparams}
\end{aligned}
\end{equation}
The cosmological parameters are the ones already considered in the
previous section, $\OmegaB$, $\OmegaCDM$, $h$, $\tau$ and $\thetaMC$,
and their priors are the same. The remaining parameters are related to
astrophysics, foregrounds and the instrumental systematics associated
with the Planck satellite. A complete description of their meaning,
and priors, can be found in \Refc{Ade:2013kta}.

\par

In order to test the robustness against prior choices, we have also
performed the same slow-roll analysis but starting from a flat prior
on $\epsstar{1}\in[0.00001,0.2]$. Such a prior implicitly favors
models producing a larger tensor-to-scalar ratio. 

In \Fig{fig:tetas},
we have represented the marginalized posteriors for all the
cosmological, astrophysics and nuisance parameters obtained from
either the Planck likelihood alone, or the Planck and BICEP2
likelihoods combined. This figure also shows these posteriors in the
case of our two prior choices on $\epsstar{1}$. All of the $\tetas$
posteriors are robust with respect to the prior choices and the
combination of data used.

In \Figs{fig:bcpsr} and \ref{fig:bcpsr_2D}, the one- and
two-dimensional posteriors in the slow-roll parameter space have been
represented for the same four combinations of prior choices and data
sets. The tension between Planck and BICEP2 is particularly visible on
the posterior for $\log(\epsstar{1})$ (Jeffreys' prior on
$\epsstar{1}$). As visible on the lower panels of \Fig{fig:bcpsr},
choosing a flat prior for $\epsstar{1}$ slightly reduces the tension
but, as explained above, would implicitly favor models having a large
tensor-to-scalar ratio. As expected, Planck and BICEP2 data together
completely determine the first two Hubble flow functions and one
obtains the two-sigma confidence intervals
\begin{equation}
0.0054 <\epsstar{1}< 0.013,\qquad 0.00013 <\epsstar{2}< 0.041,
\label{eq:eps12cl_log}
\end{equation}
for a Jeffreys' prior on $\epsstar{1}$ and
\begin{equation}
0.0056 <\epsstar{1}< 0.014, \qquad -0.0011 < \epsstar{2}<  0.039,
\label{eq:eps12cl}
\end{equation}
for a flat prior on $\epsstar{1}$. Because the spectral index is well
constrained by Planck alone (see the discussion in \sectionc{sec:pl}),
\Figs{fig:bcpsr} and \ref{fig:bcpsr_2D} show that combining Planck and
BICEP2 also induces a one-sigma shift of the posterior for the second
Hubble flow function towards vanishing values. On the contrary, the
third Hubble flow function $\epsstar{3}$ remains unconstrained and
unaffected by the inclusion of BICEP2.

In order to assess how much of these results come from the tension
between the Planck and BICEP2 data sets, we now estimate the Bayes
factor $\calRsr$ defined as the ratio between the probability of
compatibility and the probability of incompatibility.

\subsection{Compatibility for the Slow-Roll Model}
\label{sec:calRsr}

As discussed in \sectionc{sec:combining}, the compatibility between
BICEP2 and Planck can be evaluated from the Bayesian measure
\begin{equation}
\calRsr = \dfrac{\Evid(\Dp,\Db|\sr)}{\Evid(\Dp|\sr) \Evid(\Db|\sr)}\,,
\label{eq:Rsr}
\end{equation}
where $\sr$ refers to the model under scrutiny, namely
slow-roll. Here, we have not focused yet on a particular
inflationary potential as we have sampled the whole slow-roll
parameter space (in addition to the cosmological
parameters). Nonetheless, this can still be interpreted as having
chosen phenomenological inflationary priors, that we refer to as the slow-roll model, $\sr$. These priors have been mentioned earlier and are
$\ln(10^{10}\Pstar)\in[2.7,4.0]$, $\log(\epsstar{1})\in[-5,-0.7]$,
$\epsstar{2}\in[-0.2,0.2]$ and $\epsstar{3}\in[-0.2,0.2]$, plus the
standard priors for the cosmological and astrophysical parameters (see
\sectionc{sec:srpostb2}). Evaluating \Eq{eq:Rsr} requires the
computation of the three integrals given by \Eq{eq:defevidence} for
$\Dp$ (Planck), $\Db$ (BICEP2) and $\{\Dp,\Db\}$ (combined) which are
eight-dimensional for BICEP2 and twenty two-dimensional for the
others. This is technically non-trivial as evaluating the likelihood
at each point of the parameter space requires a complete integration
of the cosmological perturbations with {\CAMB}. In order to minimize
the number of likelihood evaluations and maximize convergence speed,
we have used the nested sampling algorithm as implemented in
{\MULTINEST} to estimate each evidence~\cite{Feroz:2007kg,
  Feroz:2008xx, Feroz:2011bj}. A target accuracy of $1\%$ has been
used together with a number of live points ranging from $1000$ to
$20000$, depending on the dimensionality of space. Moreover, for each
evidence, we have performed a few runs having half the number of live
points in order to estimate any systematic uncertainties. The resulting numerical estimate is
\begin{equation}
\ln(\calRsr) = -0.01 \pm 0.4,
\label{eq:Rsrnum}
\end{equation}
where the quoted error is a systematic evaluated over the various
runs. In appendix~\ref{sec:analytical}, we discuss a
semi-analytic method to calculate $\calRsr$ that requires only one
integration over the BICEP2 likelihood. The result quoted in
\Eq{eq:Rsranalytic} matches the above numerical value.

Such a value for $\calRsr$ is very close to unity and signals equal
probability of Planck and BICEP2 data to be compatible or
incompatible. Let us emphasize that, on the Jeffreys' scale, strong
compatibility would have required $\ln (\calRsr) > 5$ while strong
incompatibility would have been $\ln(\calRsr) < -5$. With values of
$|\ln(\calRsr)|<1$, we are in the inconclusive region, namely no
conclusion can be drawn on the compatibility of the two data sets. As
we illustrate in appendix~\ref{sec:toy}, the fact that we find
$|\ln(\calRsr)|<1$ is a non-trivial result. The tension between the
Planck and BICEP2 posteriors on $\epsstar{1}$ (or $r$) visible in
\Fig{fig:b2sr} ends up being compensated by the agreement between the
informative posteriors for $\Pstar$ and $\thetaMC$ (see
\Fig{fig:b2sr}). Let us stress that discussing the compatibility of
two data sets by estimating how much the likelihoods overlap in one
direction only, without specifying any prior and without marginalizing
over the other parameters, is misleading~\cite{Audren:2014cea}. As can
be seen in \Fig{fig:pl}, even after marginalization, the amount of
overlapping between the $r$-posteriors is by nature
prior-dependent. For this reason, in the following, we will discuss
the compatibility between Planck and BICEP2 by using the well defined
Bayesian measure $\calR$. In particular, even though all the {\EI}
models belong to the slow-roll class, their prior space are completely
different and their respective $\calR$ value will accordingly be
modified (see \Fig{fig:Rfactor}).

Since there is no evidence for incompatibility for the slow-roll
model, we now derive various results applicable to the slow-roll class
in general and obtained by combining Planck and BICEP2.

\subsection{Energy Scale of Inflation}
\label{sec:energyscale}

\begin{figure}
\begin{center}
\includegraphics[width=0.5\wdblefig]{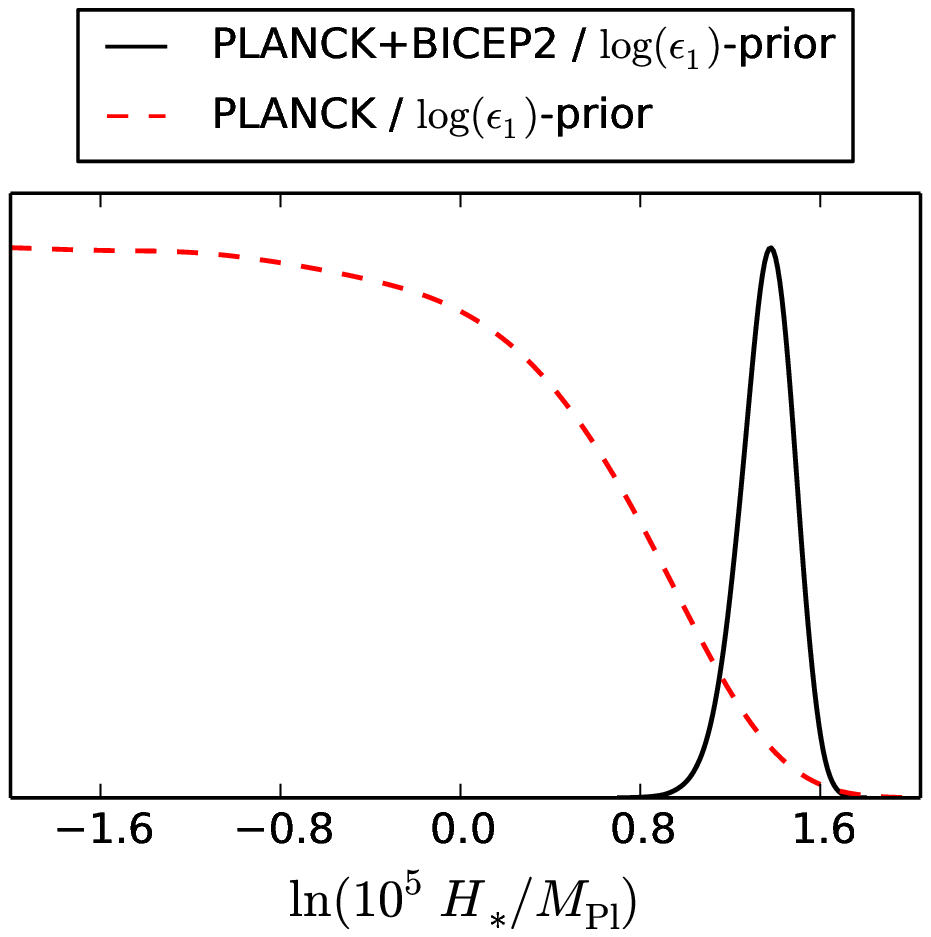}
\includegraphics[width=0.5\wdblefig]{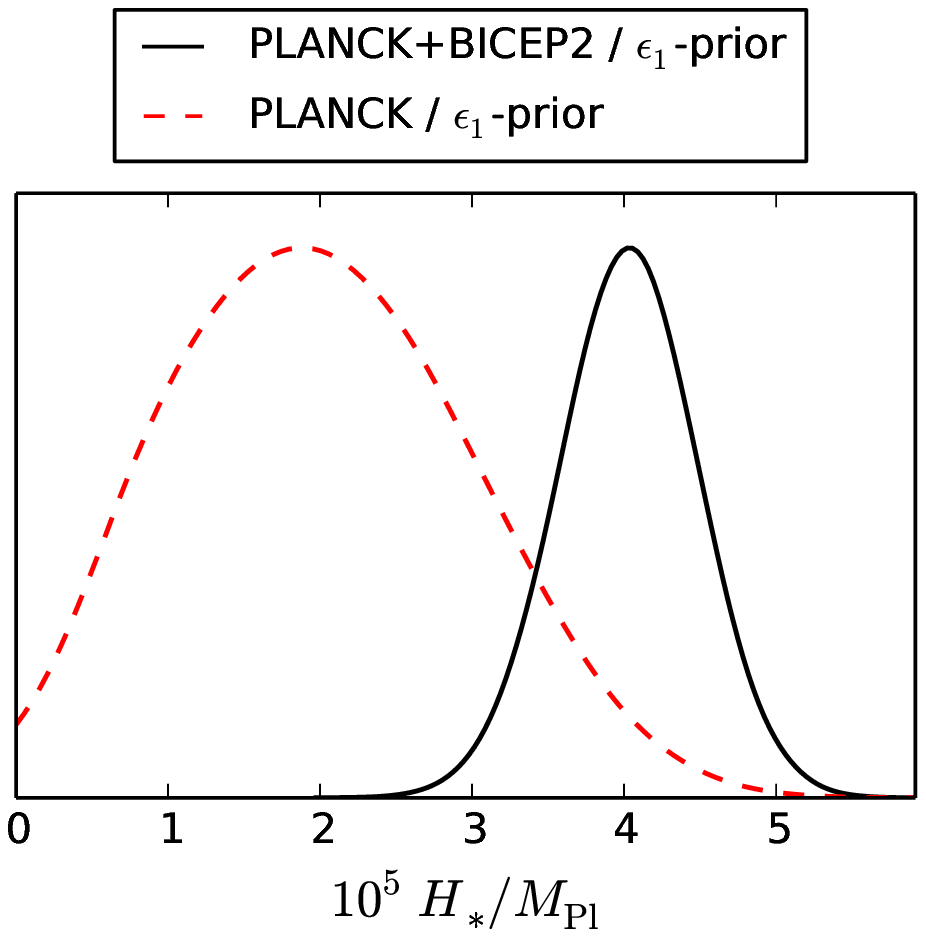}
\caption{Marginalized posterior distribution for the inflationary
  Hubble parameter at the time of pivot crossing. BICEP2 measures the
  energy scale of inflation.}
\label{fig:Hstar}
\end{center}
\end{figure}

The correct Bayesian way to determine the energy scale of inflation is
to compute the posterior distribution of the Hubble scale at the
pivot crossing time, namely for the quantity $\Hstar$ appearing in
\Eqs{eq:pzeta} and \eqref{eq:ph}. This can be done by importance
sampling from the posteriors already obtained on $\Pstar$,
$\epsstar{1}$, $\epsstar{2}$ and
$\epsstar{3}$~\cite{Lewis:2002ah}. From \Eq{eq:pzeta}, one has at
second order in slow roll
\begin{equation}
\begin{aligned}
\dfrac{\Hstar^2}{\Mpl^2} & = 8 \pi^2 \epsstar{1} \Pstar\left[1+ 2(1+C)
  \epsstar{1} + C \epsstar{2} \right],
\end{aligned}
\end{equation}
and we have plotted its posterior in \Fig{fig:Hstar}. Assuming a
Jeffreys' prior on $\epsstar{1}$, Planck and BICEP2 data combined give the
two-sigma confidence interval
\begin{equation}
1.1 < \ln\left(10^5 \dfrac{\Hstar}{\Mpl} \right) < 1.6,
\label{eq:lnH}
\end{equation}
with a mean value at $\ln(10^5 \Hstar/\Mpl)=1.36$, namely $\Hstar
\simeq 9.5 \times 10^{13}\,\GeV$. Starting from a flat prior on
$\epsstar{1}$, one obtains instead
\begin{equation}
3.1 <10^5 \dfrac{\Hstar}{\Mpl} < 4.9,
\label{eq:H}
\end{equation}
and a mean value at $10^5 \Hstar/\Mpl = 4.02$, giving $\Hstar \simeq
9.8 \times 10^{13}\,\GeV$. Those values can be converted into
gravitating energy scales through the Friedmann-Lema\^{\i}tre
equation, {\ie}
\begin{equation}
\rho_*^{1/4} = 3^{1/4} \sqrt{\Hstar \Mpl}.
\label{eq:V}
\end{equation}
One finds the corresponding values $\rho_*^{1/4} \simeq 2.00 \times
10^{16}\,\GeV$ (Jeffreys' prior on $\epsstar{1}$) and $\rho_*^{1/4}
\simeq 2.03 \times 10^{16}\,\GeV$ (flat prior on $\epsstar{1}$).

\subsection{Power Law Derived Parameters}
\label{sec:pl}

\begin{figure}
\begin{center}
\includegraphics[width=\wdblefig]{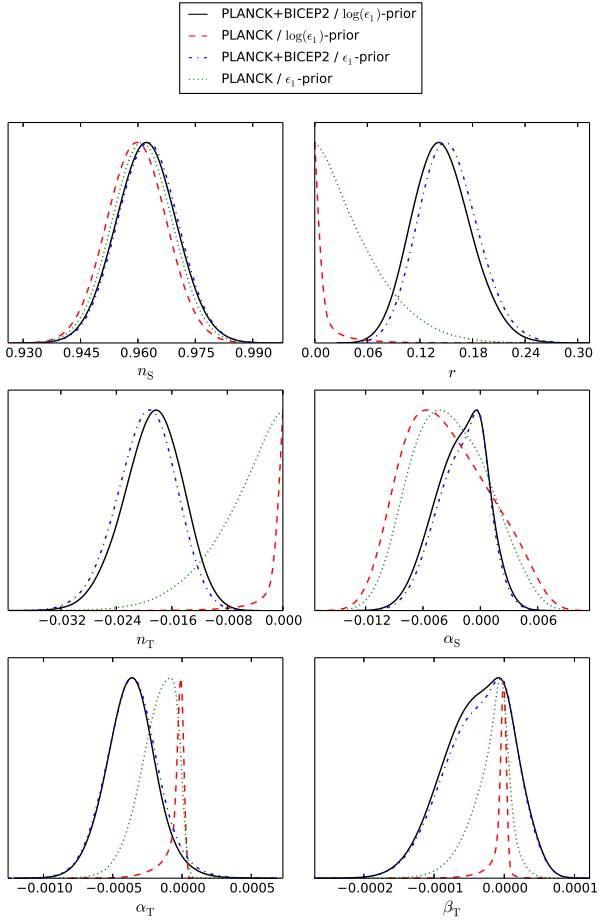}
\caption{Marginalized posterior distributions for the derived power law
  parameters $\nS$, $r$, $\nT$, $\alphaS$, $\alphaT$ and $\betaT$
  obtained by importance sampling from the second order slow-roll
  parameters. Within slow-roll inflation, the running $\alphaS$ is more tightly
  constrained when the BICEP2 data are included.}
\label{fig:pl}
\end{center}
\end{figure}

\begin{figure}
\begin{center}
\includegraphics[width=\wdblefig]{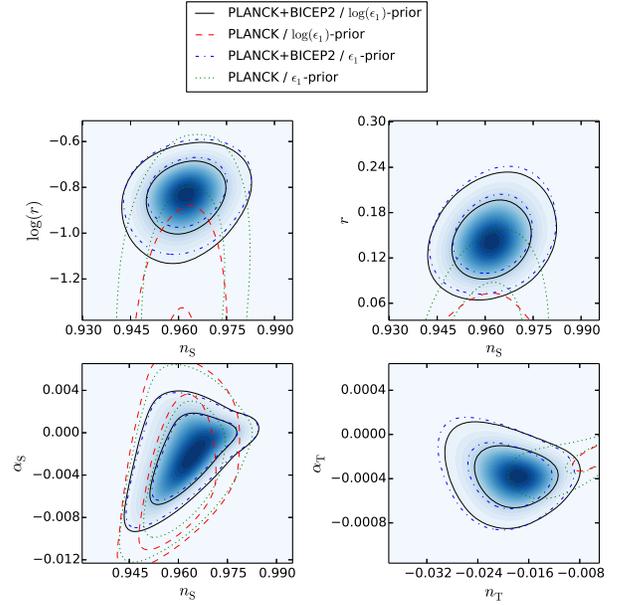}
\caption{Two-dimensional posterior distributions for some of the
  derived power law parameters. The blue shading
  density traces the mean likelihood values for Planck+BICEP2
  (Jeffreys' prior on $\epsstar{1}$).}
\label{fig:pl_2D}
\end{center}
\end{figure}

Similarly, as it is explicit from \Eqs{eq:pzeta} and \eqref{eq:ph},
the spectral indices $\nS$ and $\nT$, the tensor-to-scalar ratio $r$
and the runnings $\alphaS$ and $\alphaT$ are completely given in terms
of the Hubble flow functions. At second order in slow roll, the
spectral indices read~\cite{Martin:2013uma, Jimenez:2013xwa}
\begin{equation}
\begin{aligned}
\nS & =1 -
(2\epsstar{1} + \epsstar{2}) - 2\epsstar{1}^2 -
(3+2C)\epsstar{1}\epsstar{2} -C \epsstar{2}\epsstar{3}, \\
\nT & = -2\epsstar{1}-2\epsstar{1}^2-2(1+C)\epsstar{1}\epsstar{2},
\end{aligned}
\label{eq:nst}
\end{equation}
while the tensor-to-scalar ratio can be expressed as
\begin{equation}
r  = 16\epsstar{1} (1+C\epsstar{2}).
\label{eq:r}
\end{equation}
The scalar and tensor running are given by
\begin{equation}
\alphaS = -2\epsstar{1}\epsstar{2}-\epsstar{2}\epsstar{3},\qquad
\alphaT  =-2\epsstar{1}\epsstar{2}, 
\label{eq:alphast}
\end{equation}
respectively. Finally, let us mention that the running of the running
for the tensor mode is also completely specified by the first three
Hubble flow functions and read
\begin{equation}
\betaT = - 2 \epsstar{1} \epsstar{2} \left( \epsstar{2} + \epsstar{3} \right).
\end{equation}
At leading order in slow roll, those equations can be recast into the so-called
consistency relations
\begin{equation}
\begin{aligned}
r & \simeq -8\nT , \\
\alphaT & \simeq \frac{r}{8}\left[\frac{r}{8} +
  \left(\nS-1\right)\right], \\
\betaT & \simeq \alphaT(1-\nS) +
\dfrac{r}{8}(\alphaS - 2 \alphaT).
\end{aligned}
\label{eq:consistency}
\end{equation}
Using again importance sampling, the posterior distributions for $\nS$,
$\nT$, $r$, $\alphaS$, $\alphaT$ and $\betaT$ have been represented in
\Figs{fig:pl} and \ref{fig:pl_2D}. In particular, let us stress that,
within slow roll inflation, a running spectral index for the scalar
modes cannot help to alleviate the tension between Planck and BICEP2
data. On the contrary, we see that the posterior of $\alphaS$ is more
restricted around vanishing values by adding the BICEP2 data. As it is
explicit in \Eq{eq:alphast}, $\alphaS$ is a small quantity which is
proportional to $\epsstar{2}$, the posterior of which is being shifted
towards zero when the BICEP2 data are considered (see
\sectionc{sec:srpost}). From this equation one has
\begin{equation}
|\alphaS|_{\max} \simeq |\epsstar{2}|_{\max} \left(2|\epsstar{1}|_{\max}
  + |\epsstar{3}|_{\max} \right) \simeq |\epsstar{2}|_{\max} |\epsstar{3}|_{\max},
\label{eq:alphamax}
\end{equation}
the third Hubble flow function $\epsstar{3}$ being the largest term
since it is unconstrained [$\max{(|\epsstar{3}|)}=0.2$]. Therefore,
shifting $\epsstar{2}$ towards small values implies the same for
$\alphaS$. This effect could have been expected as Planck alone
strongly constrains the spectral index, which is given by
\Eq{eq:nst}. Increasing $\epsstar{1}$ at fixed $\nS$ imposes to
decrease $\epsstar{2}$ by twice the amount. Therefore,
\Eq{eq:alphamax} implies that the maximal values of $|\alphaS|$ will
be accordingly reduced.

From the posteriors represented in \Fig{fig:pl}, Planck and BICEP2
data combined yield the following $95\%$ confidence intervals
\begin{equation}
\begin{aligned}
0.947 < \nS < 0.978,  &\quad -0.0074 < \alphaS < 0.0025,\\
 -1.07 < \log(r)< -0.67, &\quad -0.027 < \nT < -0.011,
\end{aligned}
\end{equation}
and
\begin{equation}
\begin{aligned}
-7.1 \times 10^{-4} & <  \alphaT < -3.1 \times 10^{-6},\\
-1.3 \times 10^{-4} & <  \betaT < 5.0 \times 10^{-5},
\end{aligned}
\end{equation}
when a Jeffreys' prior is assumed on $\epsstar{1}$. These bounds are
relatively robust against the prior choices. Indeed, assuming instead a flat
prior on $\epsstar{1}$ gives
\begin{equation}
\begin{aligned}
0.947 < \nS < 0.978,  &\quad -0.0071 < \alphaS < 0.0027,\\
0.088 < r < 0.22, &\quad -0.028 < \nT < -0.011,
\end{aligned}
\end{equation}
and
\begin{equation}
\begin{aligned}
-7.1 \times 10^{-4} & <  \alphaT < 2.7 \times 10^{-5},\\
-1.3 \times 10^{-4} & <  \betaT < 5.0 \times 10^{-5}.
\end{aligned}
\end{equation}

To conclude this section, let us stress that although there is a
tension between the Planck and BICEP2 data on the tensor-to-scalar
ratio, it does not affect the posterior values of the cosmological and
astrophysics parameters, those being already strongly constrained by
the Planck data alone. Concerning the shape of the primordial power
spectra, $\nS$ remains also unaffected while the tensor-to-scalar
ratio tension induces a drastic modification of the $\epsstar{1}$
posterior distribution and a one- to two-sigma shift of the
$\epsstar{2}$ distribution compared to the Planck data
alone. Moreover, as we have just discussed, the running of the scalar
power spectrum cannot be used within slow-roll inflation to alleviate
the above-mentioned tension, precisely because it cannot take large
enough values. Solely in slow-roll violating models of inflation, such
an explanation may be relevant~\cite{Jain:2009pm, Contaldi:2014zua,
  Hazra:2014jka, Hazra:2014aea, Abazajian:2014tqa}.

Concerning the implications for inflation, BICEP2 results provide, for
the first time, a measure of the energy scale of inflation which ends
up being at GUT scale, see \Eqs{eq:H} and \eqref{eq:V}, a major result
indeed if confirmed. The mean value of $r=0.15$ is slightly lower than
what was inferred by the BICEP2 team but this is expected as we are
here considering slow-roll inflation and have added the Planck data
which disfavor larger tensor-to-scalar ratio values. As for the
evidences, one should therefore expect all models predicting a very
small tensor-to-scalar ratio to be now strongly penalized
evidence-wise. That is why, if the BICEP2 measurements stands the test of time,
this situation would be a pivotal moment for Cosmic Inflation models.

In the following, we use the multi-dimensional posterior on $\Pstar$
and $\epsstar{i}$, derived under the Jeffreys' prior on $\epsstar{1}$,
coming from the BICEP2 data as our effective likelihood $\likeinf$.

\section{Results and Discussion}
\label{sec:result}

\begin{figure*}
\begin{center}

\includegraphics[width=\wsingfig]{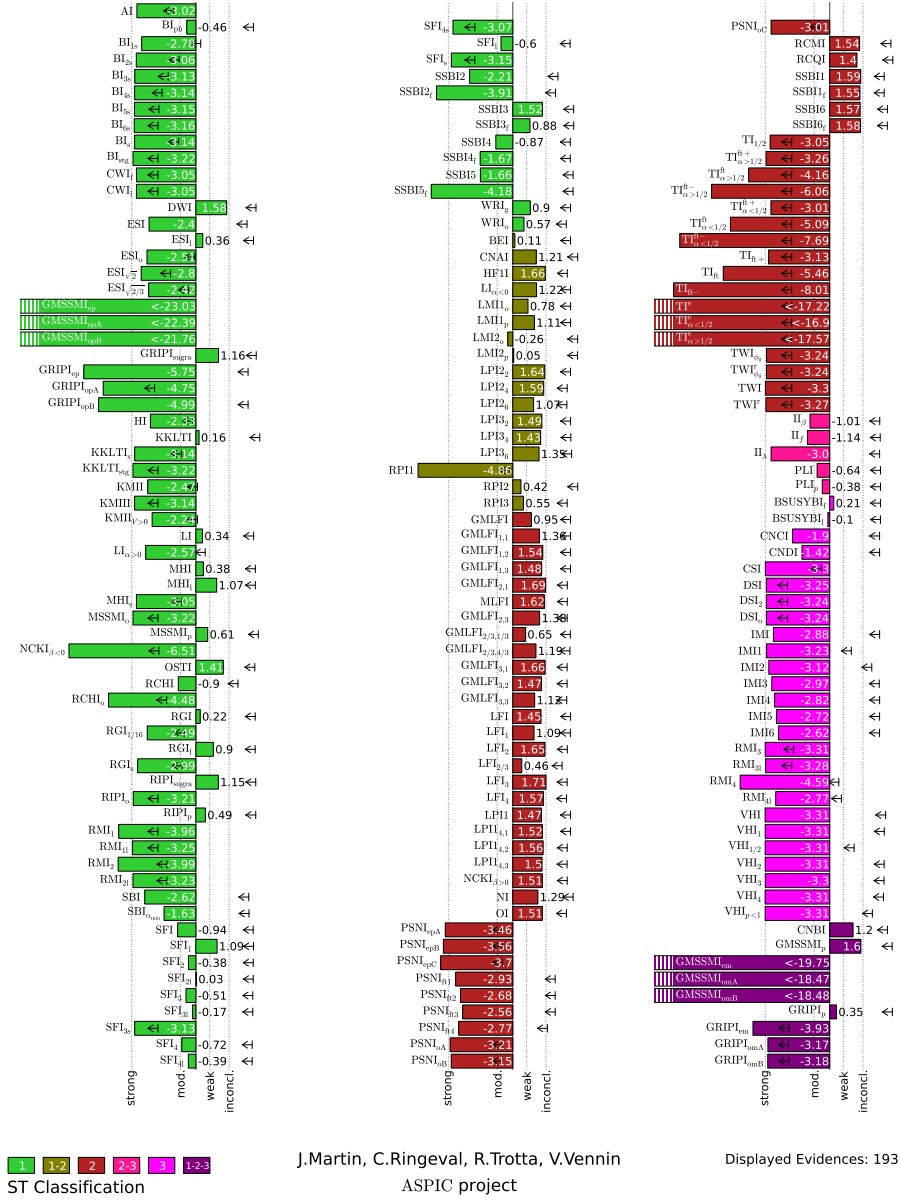}
\end{center}
\caption{Bayes factors and absolute upper bounds to the Bayes factors obtained from the BICEP2 data alone. The
  reference model is the slow-roll model ($\sr$), here viewed as a
  scenario in itself having three parameters $\ln(10^{10} \Pstar)$,
  $\log(\epsilon_1)$ and $\epsilon_2$ and whose priors are reported in
  the text. The vertical dotted lines refer to the Jeffreys' scale with
  respect to the best model, here $\lfi3$.}
\label{fig:evidBICEP}
\end{figure*}

\begin{figure*}
\begin{center}
\includegraphics[width=0.9\wdblefig]{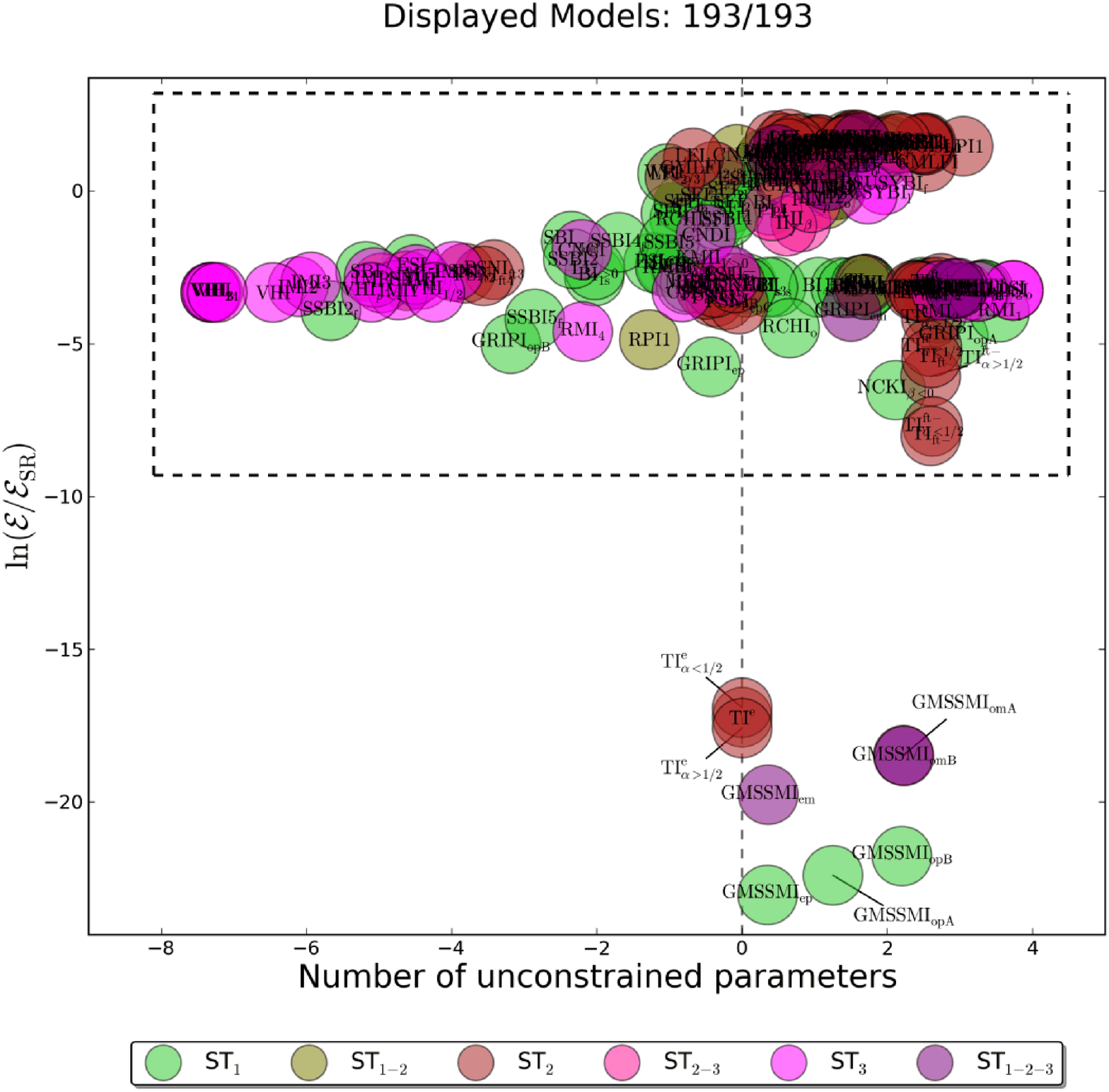}
\includegraphics[width=0.9\wdblefig]{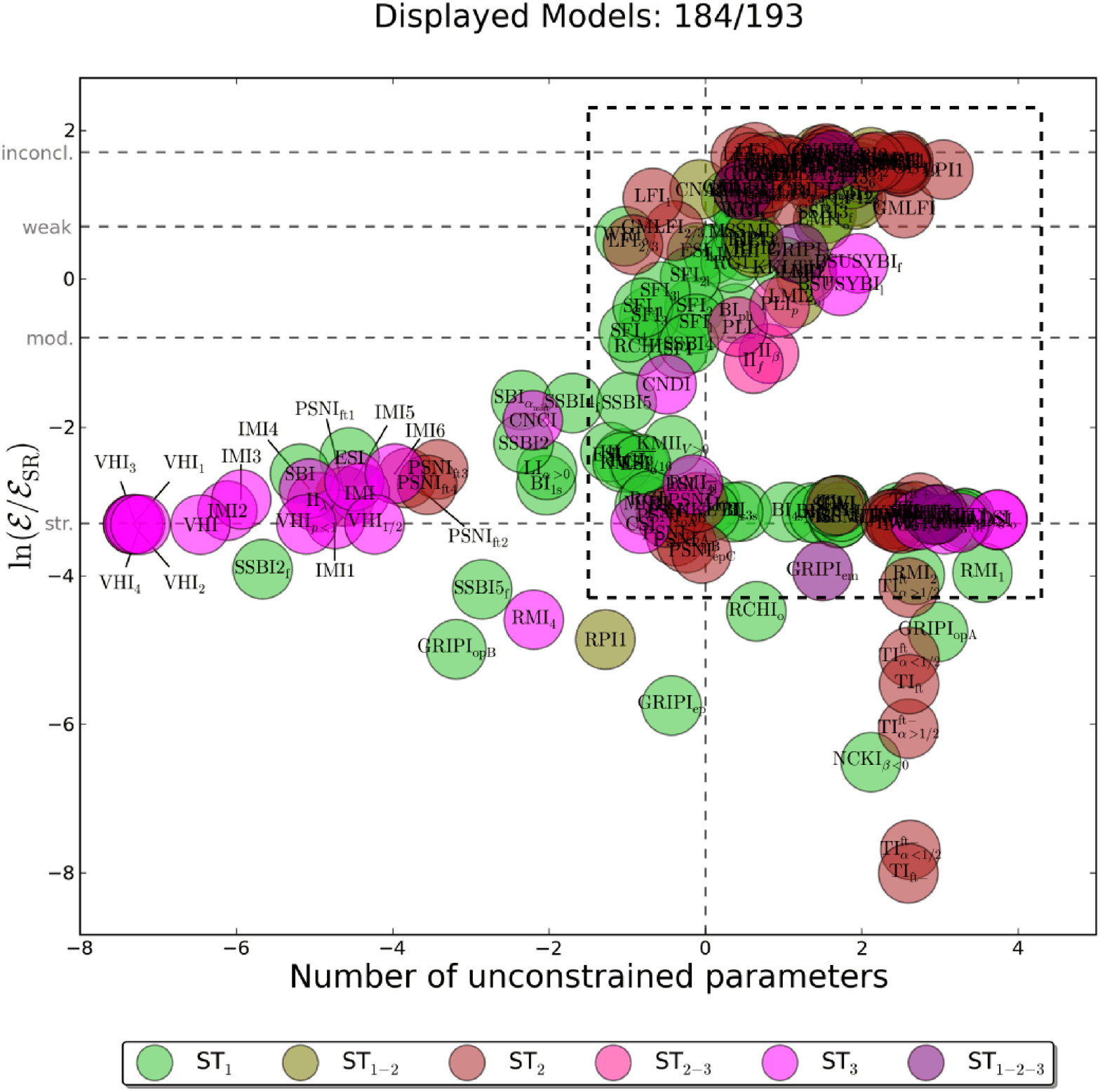}
\includegraphics[width=0.9\wdblefig]{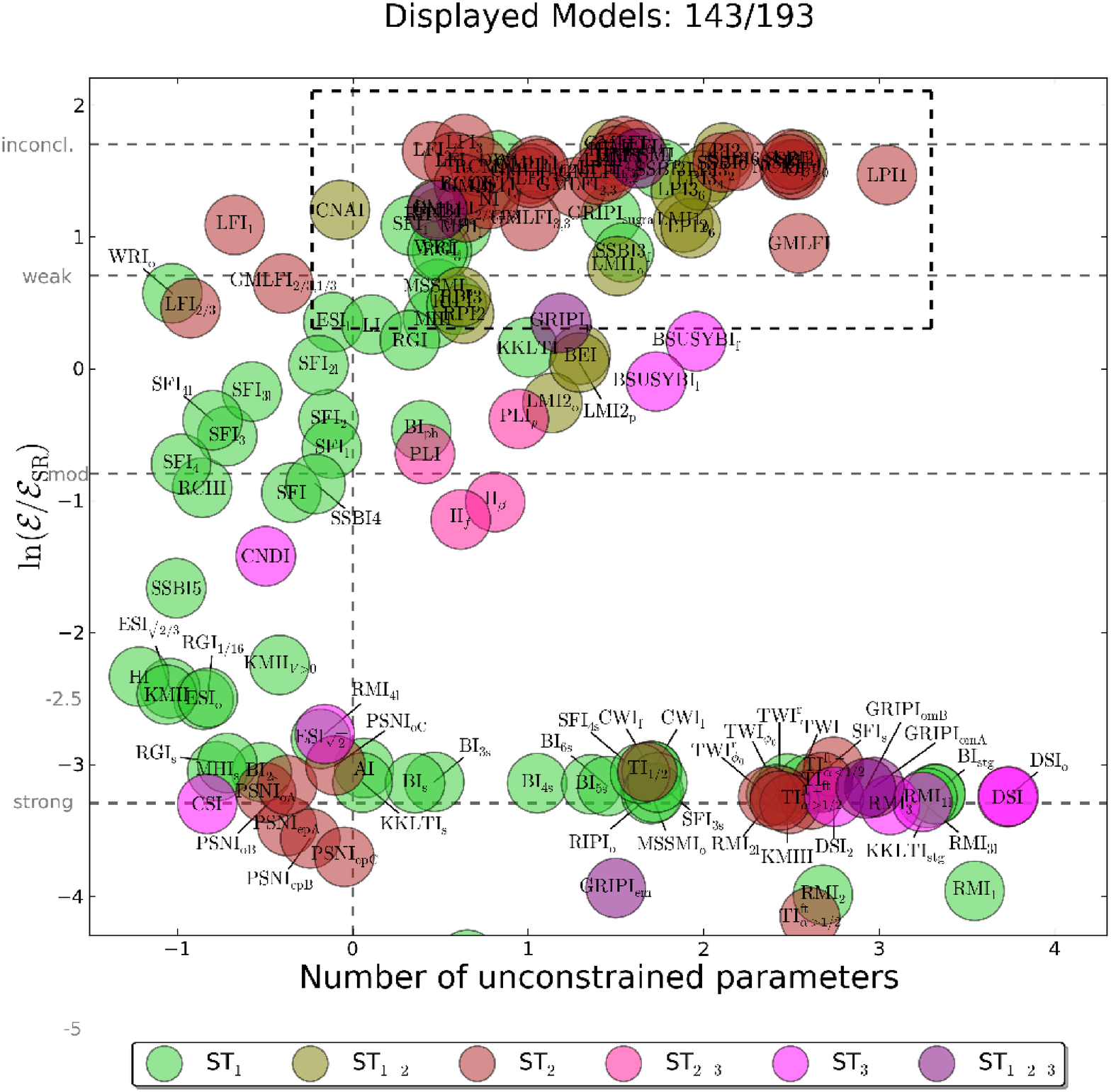}
\includegraphics[width=0.9\wdblefig]{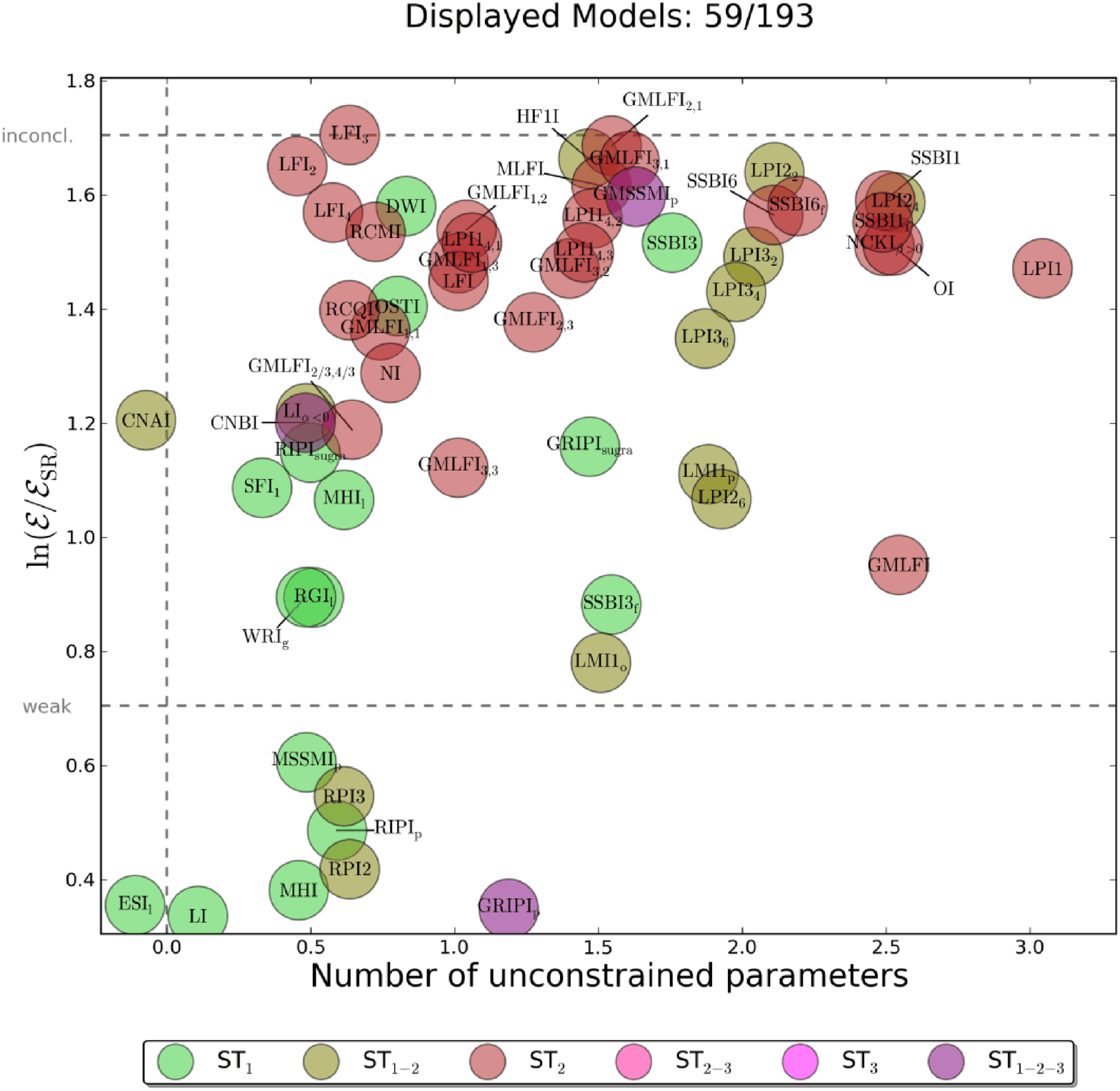}
\end{center}
\caption{Model performance assessed with both Bayesian evidence and
  number of unconstrained parameters for the BICEP2 data. The four
  panels represent different zooms in the models' space. They are to
  be read from the left to the right and from the top to the
  bottom---in this order, the black dashed rectangles frame the region
  comprised in the next panel.}
\label{fig:bayesspaceb2}
\end{figure*}

\begin{figure*}
\begin{center}
\includegraphics[width=\wsingfig]{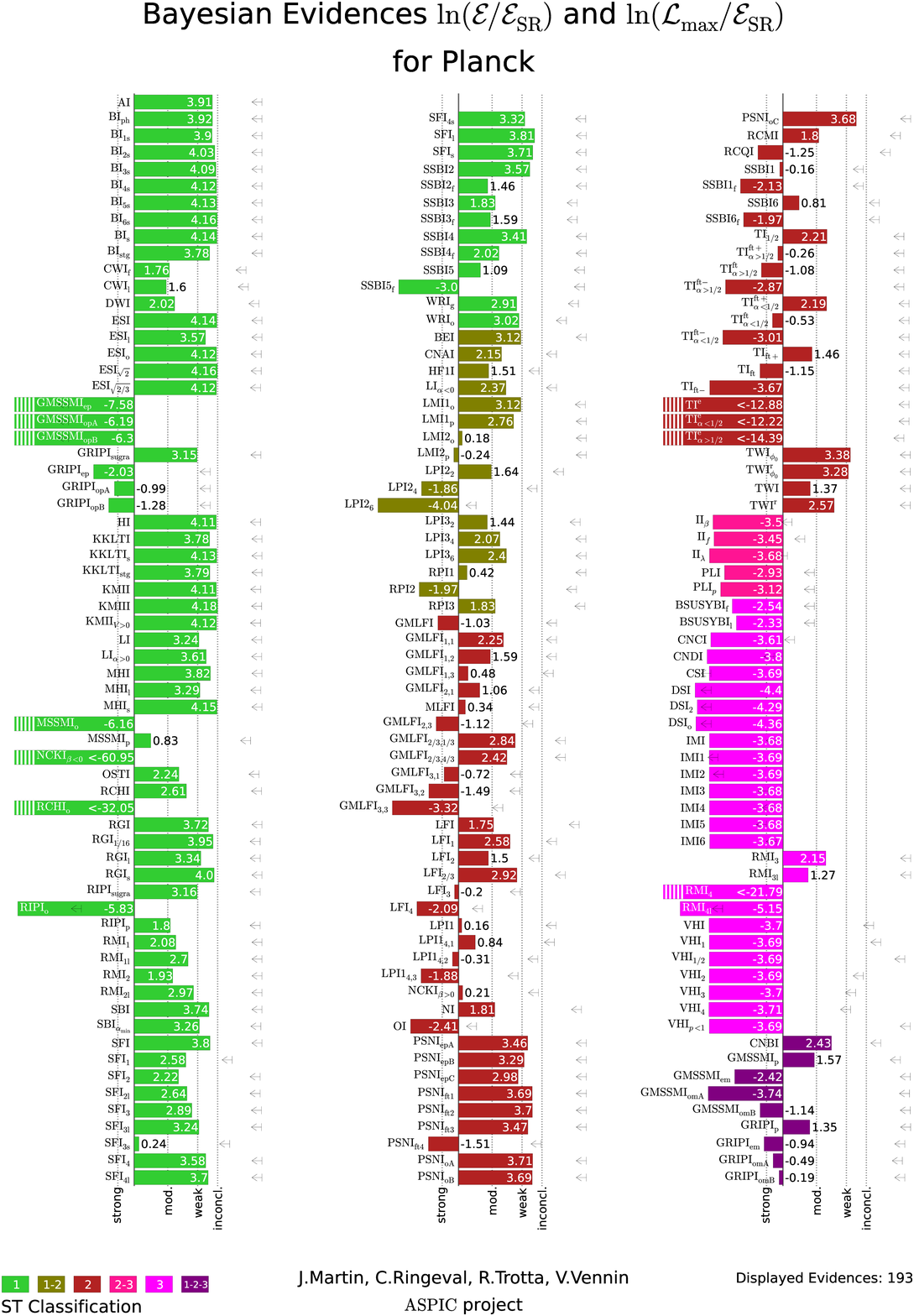}
\end{center}
\caption{Bayes factors and absolute upper bounds to the Bayes factors
  obtained from the Planck data as in Ref.~\cite{Martin:2013nzq}. The
  reference model is the same slow-roll model as in
  \Fig{fig:evidBICEP}, and the vertical dotted lines refer to the
  Jeffreys' scale with respect to the best model, here $\kmiii$.}
\label{fig:evidPlanck}
\end{figure*}

In this section, we apply the method described previously in order to
derive the Bayesian evidences and complexities for the $\Nmod=\Nevid$
models of the \EI. The complete list of models as well as a careful
discussion and justification of the priors on the free parameters
$\tetai$ can be found in Ref.~\cite{Martin:2013nzq}. In the present
article, we use the same terminology and the same choices for the
priors.

\subsection{BICEP2 Evidences}
\label{sec:evidb2}

In \Fig{fig:evidBICEP}, we show the (logarithm) of the Bayes factors,
$\Bsr{i}$, normalized to the slow-roll model, and computed with the
BICEP2 data only. The value of $\ln \Bsr{i}$ is represented by a
horizontal bar on the left if $\ln \Bsr{i}<0$ (the model $\calM_i$ is
disfavored with respect to the slow-roll model) and by a horizontal
bar on the right if $\ln \Bsr{i}>0$ (the model $\calM_i$ is favored
with respect to the slow-roll model), the length of the bar being
directly proportional to $\ln \Bsr{i}$. There is also a color code
which indicates the Schwarz Terrero-Escalante
classification~\cite{Schwarz:2004tz}. Let us briefly recall that,
according to this classification, category one (``green'' models)
corresponds to models for which the kinetic energy and the kinetic to
total energy ratio increases during inflation. Typically, this
region contains models having a plateau shape potential. As shown in
\Refc{Martin:2013nzq}, this class of models is favored by the Planck
data (see below). Category two (``red'' models) contains models for
which the kinetic energy decreases but the kinetic to total energy
ratio increases during inflation. Large field models belongs to this
region. Finally, category three (``purple'' models) refers to models
having a decreasing kinetic and kinetic to total energy
ratio. Valley hybrid inflation is an example of a model belonging to
this category; for a more detailed explanation of this classification
and its meaning, see Ref.~\cite{Martin:2013nzq}. We have also computed
the maximum value of the evidences obtained when all the parameters
have a Dirac function prior peaked at the best fit. This is indicated
by the small black arrows. They can be interpreted as upper bounds on
the evidences regardless of the priors. Finally the vertical dotted
black lines refer to the Jeffreys' scale with respect to the best
model and represent the four different categories, ``inconclusive''
(models between the first and the second vertical line, starting from
the right), ``weakly disfavored'' (between the second and the third
vertical lines), ``moderately disfavored'' (between the third and the
fourth vertical lines) and ``strongly disfavored'' (left to the fourth
vertical line), see Table~1 in Ref.~\cite{Martin:2013nzq}.

\par

As can be seen in \Fig{fig:evidBICEP}, the best model according to
BICEP2 is $\lfiTHREE$, for which $V(\phi)\propto \phi^3$. We see that
there are in fact $52$ models in the inconclusive zone (this one being
defined with respect to the best model), namely $\lfiTHREE$,
$\gmlfiTWOONE$, $\hfONEi$, $\gmlfiTHREEONE$, $\lfiTWO$, $\lpiTWOTWO$,
$\gmlfiTWOTWO$, $\gmssmi$, $\ssbiONE$, $\lpiTWOFOUR$, $\ssbiSIXf$,
$\dwi$, $\lfiFOUR$, $\ssbiSIX$, $\lpiONEFOURTWO$, $\ssbiONEf$,
$\gmlfiONETWO$, $\rcmi$, $\lpiONEFOURONE$, $\ssbiTHREE$, $\oi$,
$\nckip$, $\lpiONEFOURTHREE$, $\lpiTHREETWO$, $\gmlfiONETHREE$,
$\lpiONE$, $\gmlfiTHREETWO$, $\lfi$, $\lpiTHREEFOUR$, $\osti$,
$\rcqi$, $\gmlfiTWOTHREE$, $\gmlfiONEONE$, $\lpiTHREESIX$, $\nati$,
$\lin$, $\cnai$, $\cnbi$, $\gmlfiTWOTHREEFOURTHREE$, $\gripiS$,
$\ripiS$, $\gmlfiTHREETHREE$, $\lmiONEp$, $\lfiONE$, $\sfiONE$,
$\lpiTWOSIX$, $\mhil$, $\gmlfi$, $\wrig$, $\rgil$, $\ssbiTHREEf$ and
$\lmiONEo$, where we have ordered the list in decreasing values of the
evidences.

\par

Let us now discuss these potentials and the physical context in which
they arise. $\cnai$, $\cnbi$, $\hfONEi$, $\lmiONEo$ and $\mhil$ are
phenomenological and, therefore, difficult to embed in high energy
physics. $\lfi$ is just the general family of monomial potentials
$V(\phi)\propto \phi^p$ and, in the inconclusive zone, one finds
$\lfiTHREE$, $\lfiTWO$, $\lfiFOUR$ and $\lfiONE$. The $\gmlfi_{p,q}$
potentials (this includes $\gmlfiTWOTWO$ for which $p=q=2$) are of the
form $V(\phi)\propto (\phi/\Mp)^p\left[1+\alpha (\phi/\Mp)^q\right]$,
where $\alpha$ is a parameter controlling the amplitude of the second
term. Physically, they could represent $\lfi$ modified by some quantum
corrections~\cite{Kobayashi:2014jga, Calmet:2014lga,
  Ellis:2014rxa}. The following potentials can also be viewed as large
field corrected models: $\rcmi$, $V(\phi)\propto
(\phi/\Mp)^2\left[1-2\alpha(\phi/\Mp)^2\ln(\phi/\Mp)\right]$ and
$\rcqi$, $V(\phi)\propto (\phi/\Mp)^4[1-\alpha \ln(\phi/\Mp)]$. $\dwi$
has a $V(\phi)\propto \left[(\phi/\phizero)-1\right]^2$ which is the
sum of three monomials but was mainly used in the context of
topological inflation. In the inconclusive zone, one also finds the
SSBI potentials which are given by $V(\phi)\propto
1+\alpha(\phi/\Mp)^2+\beta (\phi/\Mp)^4$ and can be viewed as models
where the vacuum energy part of the potential is corrected by higher
order monomial terms. This is also the case for $\nckip$,
$V(\phi)\propto 1+\alpha \ln (\phi/\Mp)+\beta (\phi/\Mp)^2$ where the
corrections are also of radiative origin and loop inflation $\lin$,
$V(\phi)=\propto \left[1+\alpha \ln (\phi/\Mp)\right]$ with
$\alpha<0$. Another class of models that are among the BICEP2 winners
is LPI's, $V(\phi)\propto
(\phi/\phizero)^p[\ln(\phi/\phizero)]^q$. These scenarios are based on
super Yang-Mills theories and are also known as glue ball
inflation. Notice, however, that the value of $\phizero$ must be
super-Planckian. The model $\oi$, $V(\phi)\propto
(\phi/\phizero)^4[\ln^2(\phi/\phizero)-\alpha]$, possesses a similar
potential as well as $\osti$, $V(\phi)\propto (\phi/\phizero)^2\ln
\left[(\phi/\phizero)^2\right]$. This last scenario is physically well
motivated in the context of string theory. Unfortunately, it is used
outside its natural domain of validity since Ref.~\cite{Kofman:2002rh}
showed that it has severe problems in matching the amplitude of the
CMB anisotropies. In the inconclusive zone, one also finds inflection
point models such as $\gmssmi$, $\ripiS$ and $\gripiS$. Some of them
are also used in a non physical region. For instance, this is the case
for $\gmssmi$, $V(\phi)\propto
(\phi/\phizero)^2-2\alpha/3(\phi/\phizero)^6+\alpha/5(\phi/\phizero)^5$. The
model is based on the MSSM (Minimal Supersymmetric Standard Model)
where the inflaton field evolves along a flat direction and is,
therefore, well justified from a high energy point of view. However,
in order to be a satisfactory inflationary model, $\phizero$ must have
a \vev that is outside the natural MSSM values. $\rgil$ refers to
radion gauge inflation and has a potential given by $V(\phi)\propto
(\phi/\Mp)^2/[\alpha+(\phi/\Mp)^2]$. $\sfiONE$ is nothing but small
field inflation, $V(\phi)\propto 1-(\phi/\mu)^p$, with $p=1$. Finally,
natural inflation ($\nati$), for which $V(\phi)\propto 1+\cos(\phi/f)$
is also a good model but must be used in a domain where the scale $f$
is super-Planckian.

\par

We have also computed the Bayesian complexities, see
\Eq{eq:defcomplex}, for all the \EI models, so that the
performance of a model can be described by two numbers, its
evidence and its complexity or, equivalently see \Eq{eq:defuparam}, its
evidence and its number of unconstrained parameters. We have
represented the corresponding result in the space $\left[\Nuc,
  \ln\left(\calE/\calE_{\mathrm{best}}\right) \right]$ in
\Fig{fig:bayesspaceb2}. If one restricts oneself to models in
the ``inconclusive zone'' with a minimal number of unconstrained
parameters, \ie $0<\Nuc_{i}<1$, then one finds only $17$ models,
namely: $\lfiTHREE$, $\lfiTWO$, $\dwi$, $\lfiFOUR$, $\rcmi$, $\osti$,
$\rcqi$, $\gmlfiONEONE$, $\nati$, $\lin$, $\cnbi$,
$\gmlfiTWOTHREEFOURTHREE$, $\ripiS$, $\sfi1$, $\mhil$, $\wrig$ and
$\rgil$. It is interesting to notice that the $V(\phi)=m^2\phi^2/2$
model (\ie $\lfiTWO$) is among the models favored by BICEP2 but is not
the only one. At this stage, it would therefore be unjustified to
focus model building efforts on this scenario only.

\par

In order to compare the above results with what has been obtained by
Planck, we have reproduced in Fig.~\ref{fig:evidPlanck} the values of
the evidences, normalized to the slow-roll model, obtained with the
Planck data in Ref.~\cite{Martin:2013nzq}. This figure is identical to
Fig.~2 of Ref.~\cite{Martin:2013nzq} except that the reference model
is now different (being $\hi$ in~\cite{Martin:2013nzq}). The best Planck model
is $\kmiii$ and the $52$ models that end up being in the inconclusive
zone (with respect to the best) are: $\kmiii$, $\esisqrtTWO$,
$\biSIXs$, $\mhis$, $\bis$, $\esi$, $\biFIVEs$, $\kkltis$, $\kmiivp$,
$\biFOURs$, $\esio$, $\esisqrtTWOTHREE$, $\kmii$, $\hi$, $\biTHREEs$,
$\biTWOs$, $\rgis$, $\rgiONEONESIX$, $\bi$, $\ai$, $\biONEs$, $\mhi$,
$\sfil$, $\sfi$, $\kkltistg$, $\bistg$, $\kklti$, $\sbi$, $\rgi$,
$\sfis$, $\psnioA$, $\sfiFOURl$, $\psniftTWO$, $\psnioB$,
$\psniftONE$, $\psnioC$, $\lip$, $\sfiFOUR$, $\esil$, $\ssbiTWO$,
$\psniftTHREE$, $\psniepA$, $\ssbiFOUR$, $\twiAONE$, $\rgil$,
$\sfiFOURs$, $\mhil$, $\psniepB$, $\twiATWO$, $\sbialphamin$, $\li$,
$\sfiTHREEl$.

\par

Two remarks are in order here. Firstly, the number of models favored
is exactly the same for BICEP2 and Planck, namely $52$. This probably
illustrates the fact that $r$ is an observable which is able to
discriminate among the inflationary models much more efficiently that
$\nS$. Indeed, as above-mentioned, we have used only 4 bandpowers for
the BICEP2 data, and this already singles out the same number of
scenarios in the inconclusive zone. Secondly, there are only two models
belonging to the two lists: $\mhil$ and $\rgil$. In particular, the
fact that the Starobinsky (or Higgs) inflationary model was among the
winners according to Planck is not recovered by BICEP2. On the
contrary, this one becomes (almost) strongly disfavored compared to
$\lfiTHREE$.

In \Refc{Martin:2013nzq}, the complexity was also calculated, see
Fig.~3 of that article. We found that, among the models in the Planck
inconclusive zone, those with a minimal number of unconstrained
parameters, \ie $0<\Nuc_{i}<1$, are: $\esisqrtTWO$,
$\esisqrtTWOTHREE$, $\hi$, $\biTWOs$, $\rgis$, $\ai$, $\biONEs$,
$\mhi$, $\rgi$, $\sfiFOURl$, $\lip$, $\sfiFOUR$, $\esil$, $\rgil$,
$\mhil$, $\sbialphamin$ and $\sfiTHREEl$. Again, two models remain in
the two lists, the same as above, namely $\mhil$ and $\rgil$.

\begin{figure}
\begin{center}
\includegraphics[width=\wdblefig]{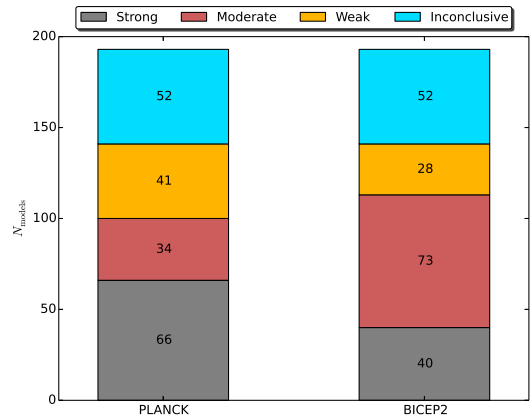}
\end{center}
\caption{Number of models within each Jeffreys' category (with respect
  to the best model) for Planck data alone and BICEP2 data alone.}
\label{fig:histo}
\end{figure}

Finally, we can summarize the data constraining power in an histogram
for the four Jeffreys' categories as represented in
Fig.~\ref{fig:histo}. We have also represented the same histogram
obtained from the Planck data. Noticing again that the BICEP2 data
used here consist only of four bandpowers for $E$ and $B$-modes, this
plot illustrates the power of measuring $r$ for inflationary
physics. However, as discussed earlier, the models lying into these
four categories weakly overlap between Planck and BICEP2 thereby
showing some tension between the data sets. Since compatibility
between data sets is a model dependent statement, we now move on to
the determination of the $\calR$ factors for all the {\EI} models.

\subsection{Compatibility of Planck and BICEP2}
\label{sec:RPb2}

\begin{figure*}
\begin{center}
\includegraphics[width=\wsingfig]{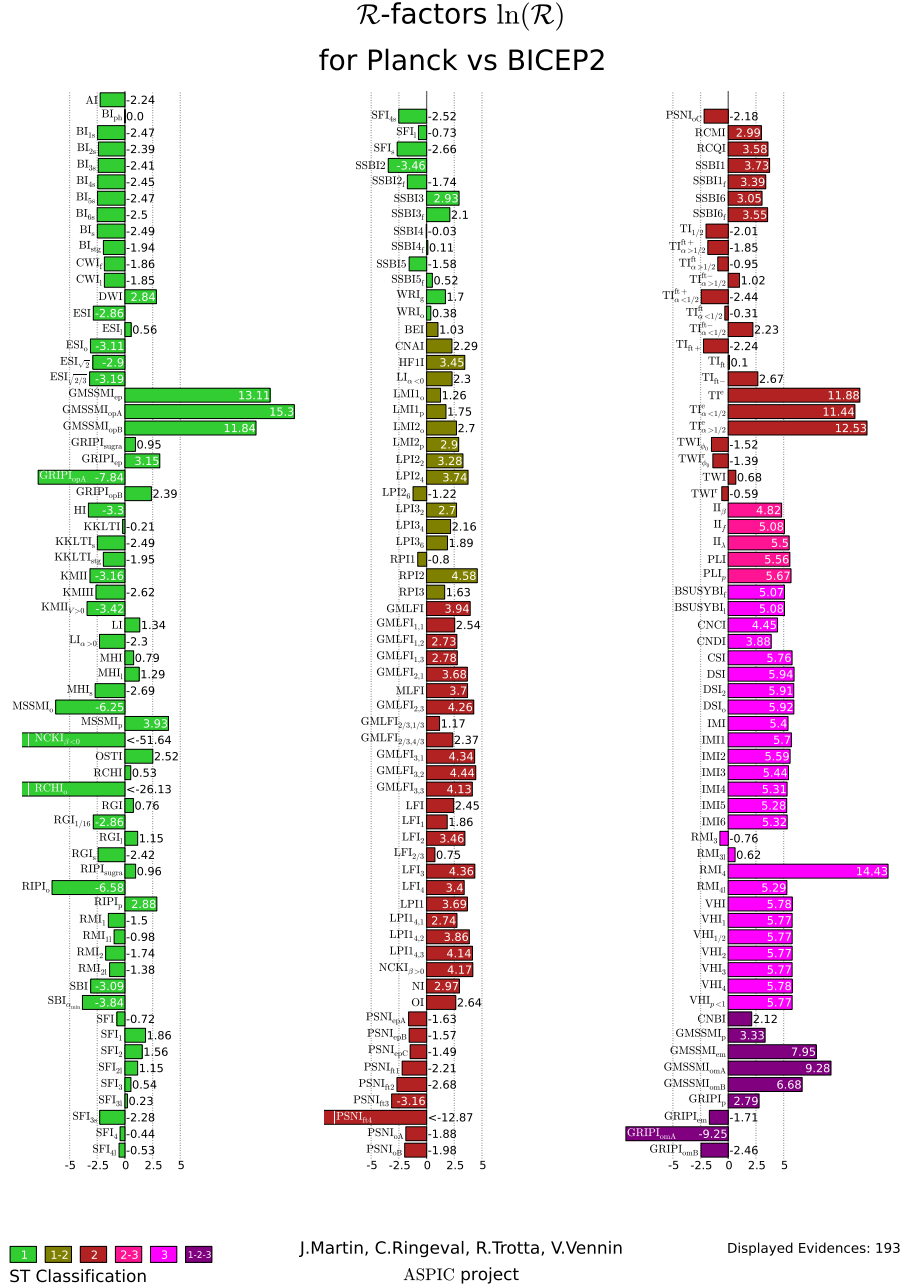}
\end{center}
\caption{Compatibility between the Planck and BICEP2 data for each
  model as measured by the $\calR$-factors.
  Positive values correspond to compatibility, negative
  values to incompatibility.}
\label{fig:Rfactor}
\end{figure*}

\begin{figure}
\begin{center}
\includegraphics[width=0.9\wdblefig]{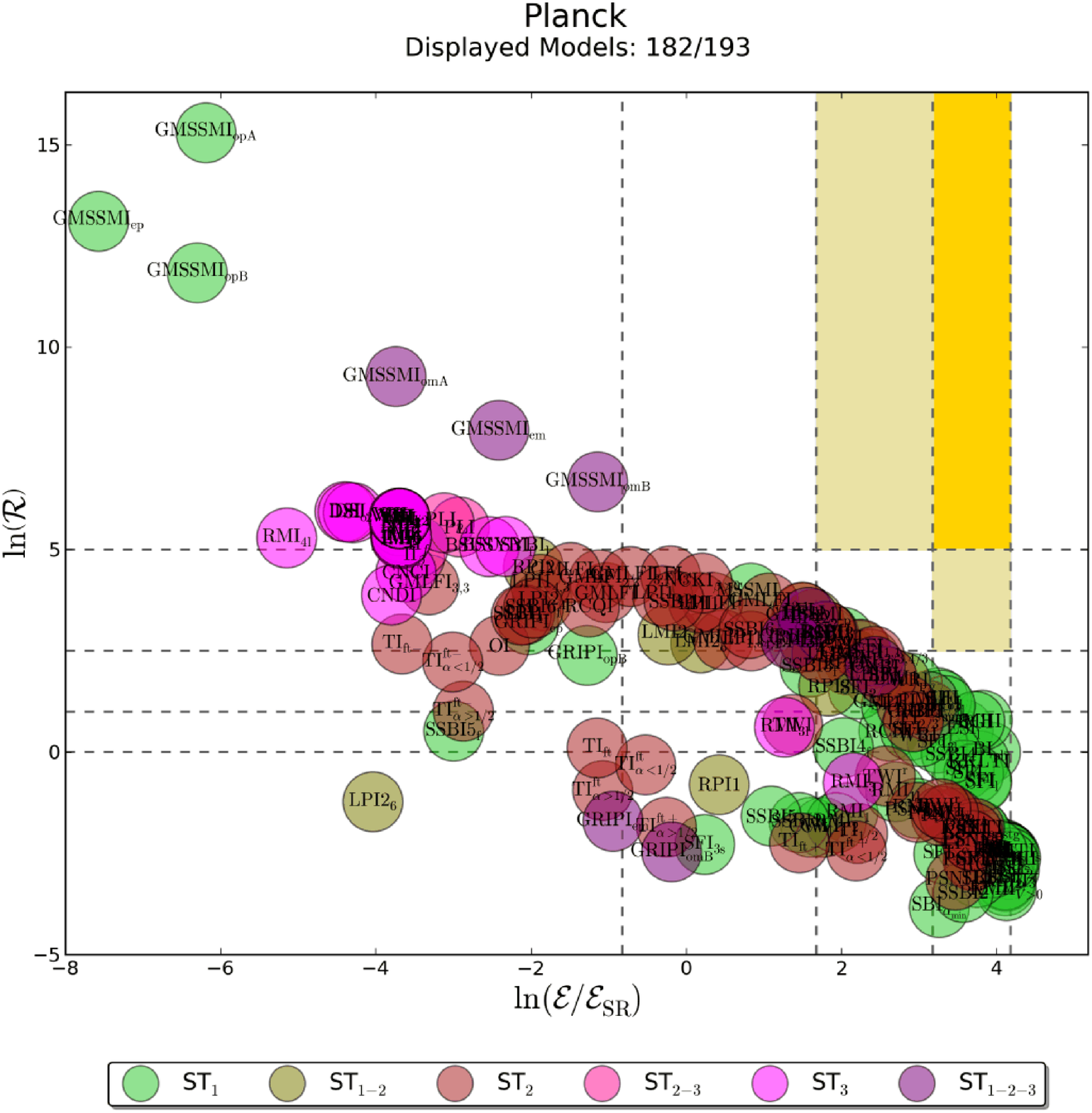}
\includegraphics[width=0.9\wdblefig]{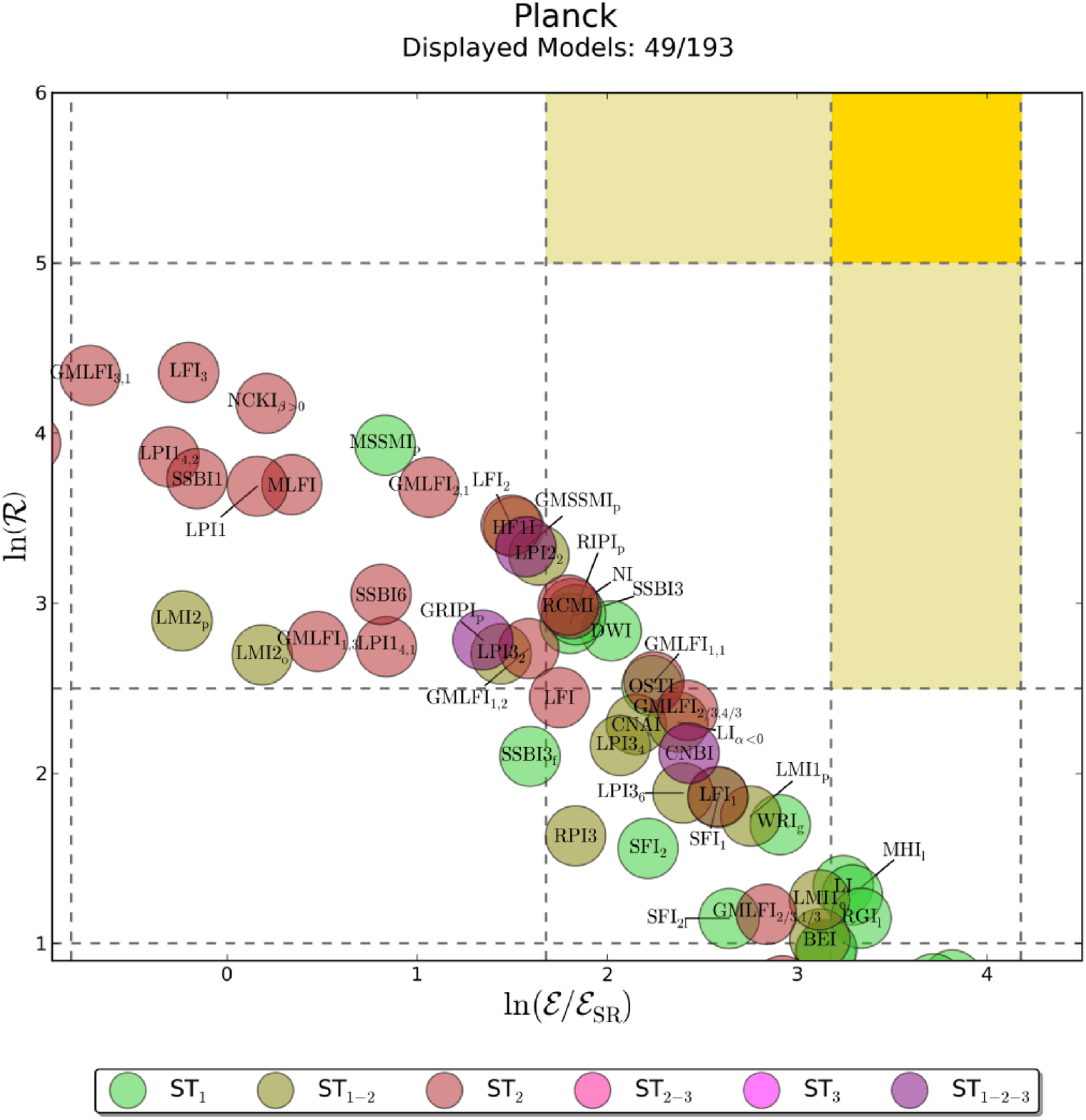}
\end{center}
\caption{Planck+BICEP2 compatibility measure, $\calR$, versus Planck's
  evidences normalized to slow roll. The yellow rectangle in the top
  right encompasses the ``strongly compatible'' models that lie in the
  Planck-alone ``inconclusive'' zone (with respect to Planck's best
  model); the light yellow rectangles encompass the ``strongly
  compatible'' models that lie in the ``weakly disfavored'' zone (top
  left) and the ``moderately compatible'' models that lie in the
  ``inconclusive'' zone (bottom right). One can see that these
  rectangles are empty. The bottom panel is a zoom into the
  neighborhood of these regions. Among the models favored by Planck
  data alone, there are only a few for which Planck and BICEP2 data
  are, at most, weakly compatible [$1<\ln(\calR) <2.5$].}
\label{fig:Rplc}
\end{figure}

\begin{figure}
\begin{center}
\includegraphics[width=0.9\wdblefig]{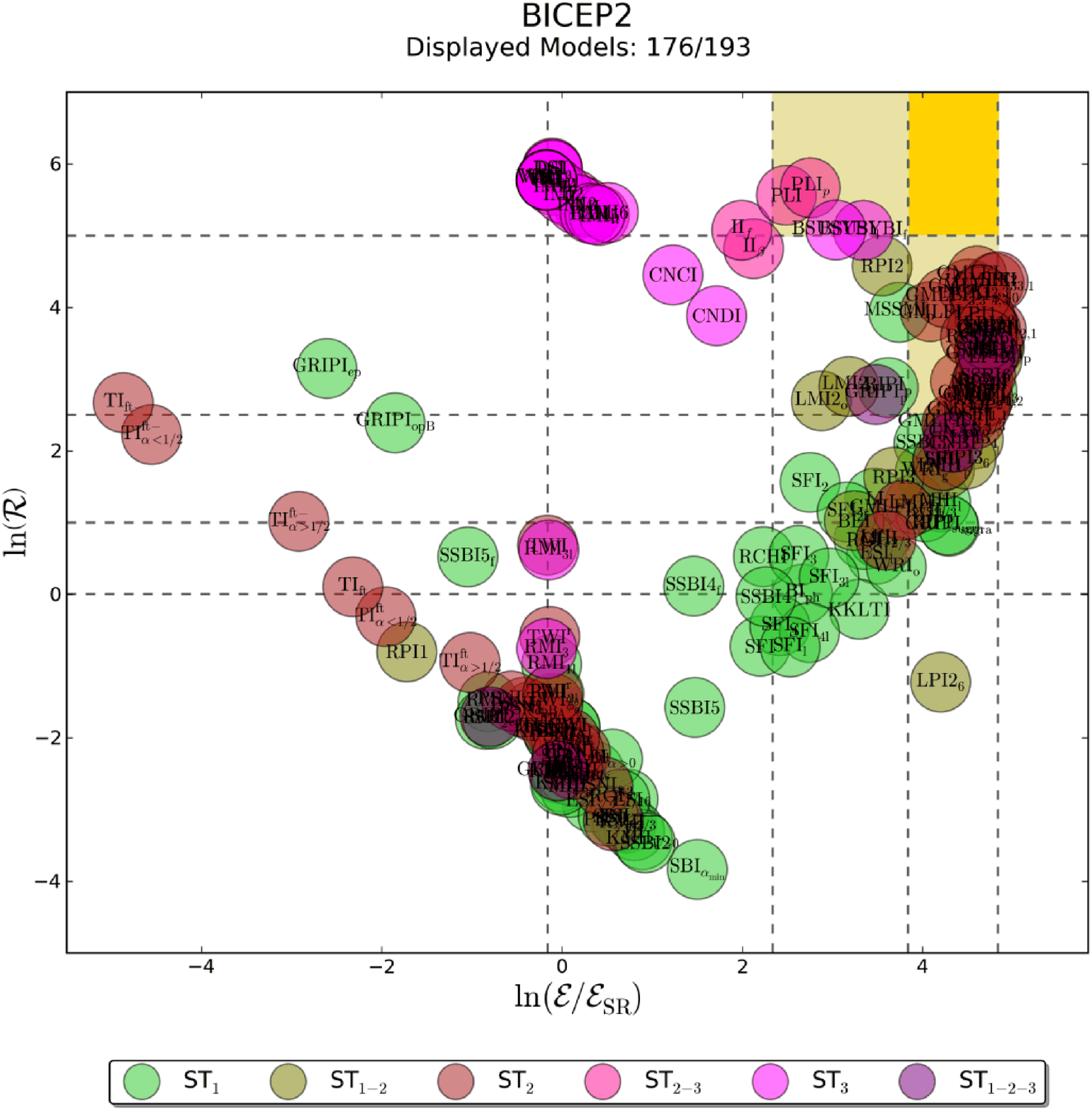}
\includegraphics[width=0.9\wdblefig]{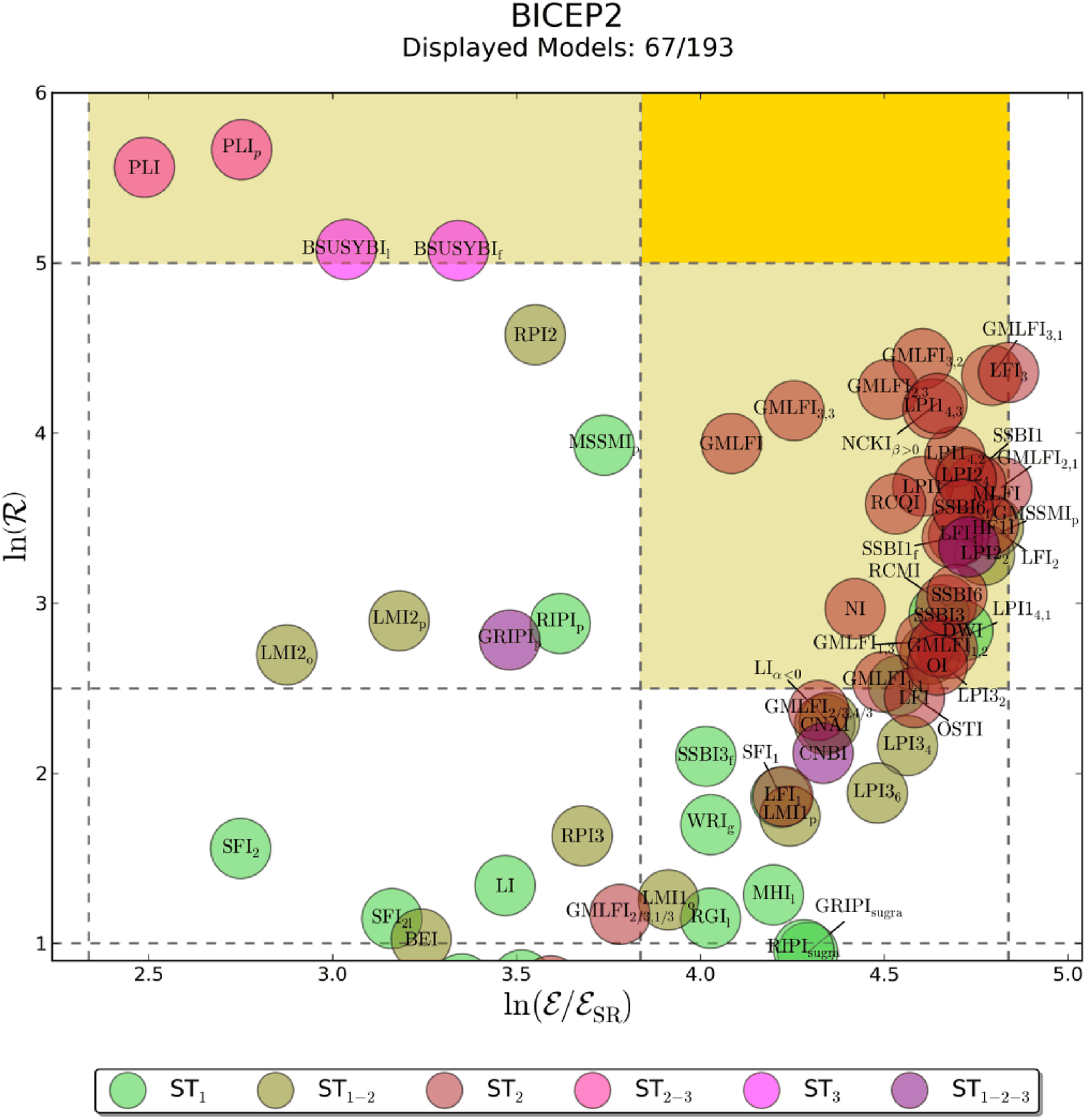}
\end{center}
\caption{Planck+BICEP2 compatibility measure, $\calR$, versus BICEP2
  evidences normalized to slow roll. The yellow rectangle in the top
  right encompasses the ``strongly compatible'' models that lie in the
  BICEP2-alone ``inconclusive'' zone (with respect to BICEP2's best
  model); the light yellow rectangles encompass the ``strongly
  compatible'' models that lie in the ``weakly disfavored'' zone (top
  left) and the ``moderately compatible'' models that lie in the
  ``inconclusive'' zone (bottom right).  The bottom panel is a zoom
  into the neighborhood of these regions. The models favored by BICEP2
  data alone are found in the region where Planck and BICEP2 are
  moderately compatible.}
\label{fig:Rbcp}
\end{figure}

\begin{figure*}
\begin{center}
\includegraphics[width=\wsingfig]{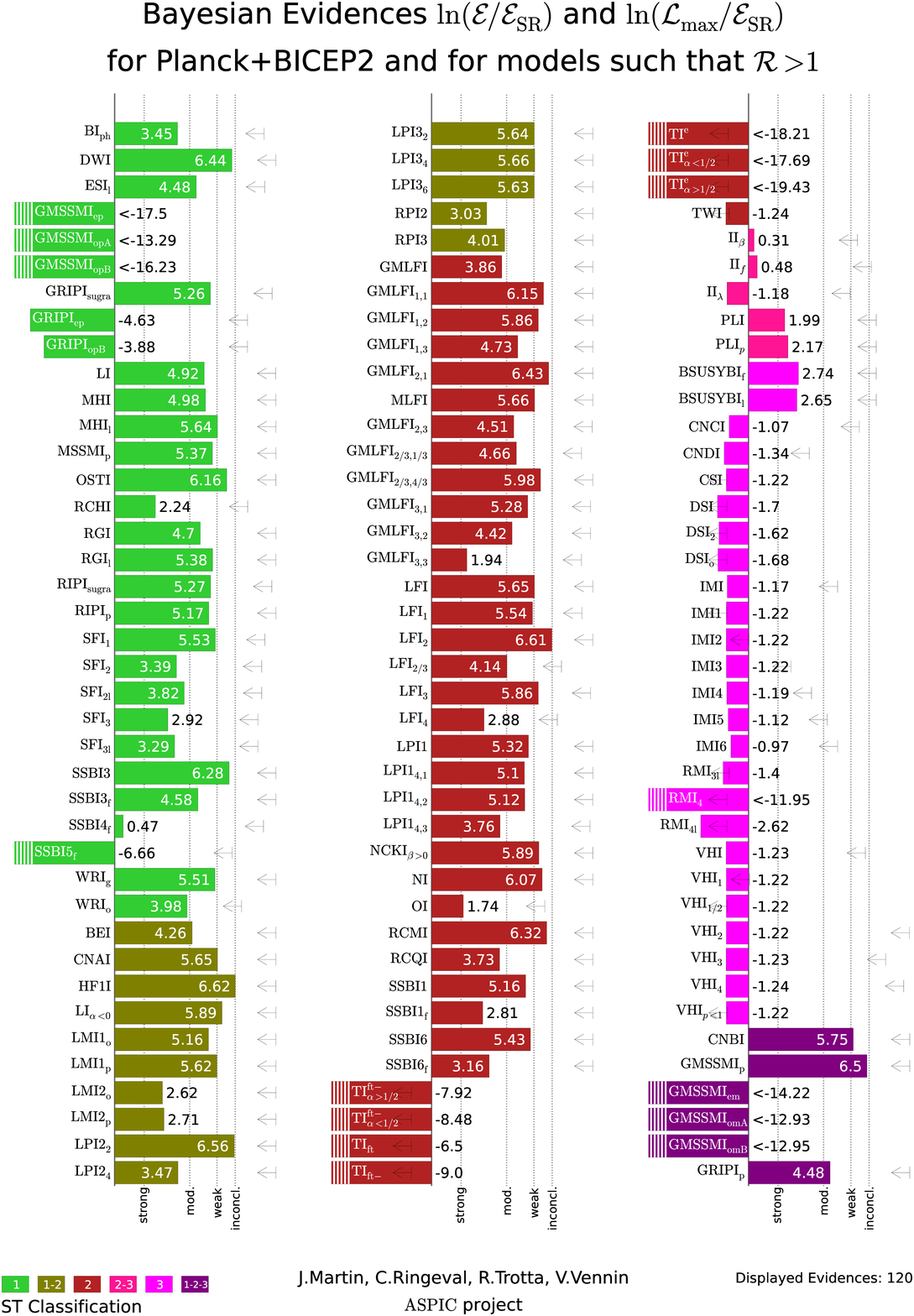}
\end{center}
\caption{Evidences (Bayes factor) and absolute upper bounds to the
  Bayes factors, from Planck and BICEP2 data combined, for the models
  such that $\calR>1$ only. The reference model is the same slow-roll
  model as in \Figs{fig:evidBICEP} and~\ref{fig:evidPlanck}, and the
  vertical dotted lines refer to the Jeffreys' scale with respect to
  the best model, here $\hfONEi$.}
\label{fig:evidPlanckBICEP}
\end{figure*}

In Fig.~\ref{fig:Rfactor} we have represented the values of $\ln
(\calR)$ for all the {\EI} models. These have been obtained using the
fast likelihood method described in \sectionc{sec:method}.  In this
plot, one notices that the data sets are compatible with certainty
[\ie at the ``strong'' level, $\ln (\calR)>5$] for $36$ models
only. They are: $\gmssmiopA$, $\gmssmiopB$, $\gmssmiep$, $\tie$,
$\tiem$, $\tiep$, $\ii$, $\iif$, $\iilambda$, $\pli$, $\plip$,
$\bsusybif$, $\bsusybil$, $\csi$, $\dsi$, $\dsiTWO$, $\dsio$, $\imi$,
$\imiONE$, $\imiTWO$, $\imiTHREE$, $\imiFOUR$, $\imiFIVE$, $\imiSIX$,
$\rmiFOUR$, $\rmiFOURl$, $\vhi$, $\vhiONE$, $\vhiONETWO$, $\vhiTWO$,
$\vhiTHREE$, $\vhiFOUR$, $\vhilONE$, $\gmssmiem$, $\gmssmiomA$, and
$\gmssmiomB$. As one can check in \Figs{fig:evidBICEP} and
\ref{fig:evidPlanck}, these models are disfavored by both Planck and
BICEP2 separately; the ones exhibiting maximum compatibility are even
ruled out. This is not surprising as $\calR$ is a combined measure of
both the reduction of prior volume brought about by the likelihood as
well as their overlap (see appendix~\ref{sec:toy}). The statistical
interpretation of these results is that both data sets agree in
disfavoring those models.

On the other hand, one may be more interested in answering the
question whether the data sets are compatible assuming the best
Planck's scenarios. In \Fig{fig:Rplc}, we have represented the same
$\calR$-factors of \Fig{fig:Rfactor} plotted against the Bayes factor
derived from the Planck data alone (the ones of
\Fig{fig:evidPlanck}). The shaded rectangles (yellow) trace the
overlapping regions of maximal evidence and maximal compatibility over
two units in the Jeffreys' scale: inconclusive plus weak zones along
the evidence direction and strong plus moderate zones along the
compatibility direction. There is no model in these regions showing
that, insofar the best inflationary models from Planck data alone are
concerned, the two data sets are in tension. In fact, only a weak
compatibility is reached for models which are already weakly
disfavored by the Planck data alone. Many of these models belong to
the ones listed earlier that were favored by BICEP2 alone (as $\nati$,
$\ssbiTHREE$, $\ripip$ \dots).

For Planck best models, the BICEP2 data cannot be brought into
compatibility with Planck, and hence the two data sets cannot be
combined to obtain meaningful updated inferences on these
scenarios. In particular, this is the conclusion for Starobinsky
inflation ($\hi$) and, therefore, it is premature to conclude about
its viability before compatibility is addressed. As we have just
showed, both data sets can be meaningfully combined only if one
focuses on scenarios which are, at least, weakly disfavored by Planck.

It is also informative to assess the compatibility of the two data sets
from the perspective of the BICEP2 best models. In \Fig{fig:Rbcp} we
have plotted the analogous of \Fig{fig:Rplc} for BICEP2, namely
the $\calR$-factors against the Bayes factors obtained from BICEP2
data alone (see \Fig{fig:evidBICEP}). The BICEP2 best scenarios now
spread into the region of moderate compatibility although there is
again no model in the strong compatibility region. Nonetheless, for
the BICEP2 best scenarios, Planck and BICEP2 data can be combined to
get more information for these scenarios.

In the light of the above considerations, in \Fig{fig:evidPlanckBICEP}
we have represented the Bayes factors obtained by combining Planck and
BICEP2 together, but only for models having $\calR> 1$ since combining
models with $\calR<1$ is meaningless.  Our chosen threshold of $\calR$
is conservative (i.e., we are not requiring $\calR \gg 1$), and it
includes scenarios under which, in the present situation, one cannot
conclude about compatibility according to the Jeffreys' scale (\ie
models having $0<\ln \calR \ll 5$). The two best models are now
$\hfONEi$ and $\lfiTWO$. Then, in the inconclusive zone (with respect
to the new best model) one has $\lpiTWOTWO$, $\gmssmi$, $\dwi$,
$\gmlfiTWOONE$, $\rcmi$, $\ssbiTHREE$, $\osti$, $\gmlfiONEONE$,
$\nati$, $\gmlfiTWOTHREEFOURTHREE$, $\lin$, $\nckip$, $\gmlfiONETWO$,
$\lfiTHREE$, $\cnbi$, $\lpiTHREEFOUR$, $\gmlfiTWOTWO$, $\cnai$,
$\lfi$, $\mhil$, $\lpiTHREETWO$, $\lpiTHREESIX$, $\lmiONEp$, $\lfiONE$
and $\sfiONE$. It is interesting to notice that, among the previous
models, none of them is in the strongly compatible zone. This is yet
another consequence of the tension between Planck and BICEP2 under an
inflationary prior assumption. We also notice that all the models in
the Planck+BICEP2 inconclusive zone are in the BICEP2 inconclusive
zone while only one (\ie $\mhil$) is in the Planck inconclusive
zone. On the other hand, the $\lfiFOUR$ scenario which was in the list
of inconclusive models for BICEP2 becomes moderately disfavored when
adding Planck.

\section{Conclusions}
\label{sec:conclusion}

Let us now summarize our main conclusions. If the BICEP2 data stand
the test of time and are confirmed as a signature of tensor modes of
inflationary origin, they do represent a major advance in our
understanding of inflation and primordial cosmology. Indeed, for the
first time, we would now have a measurement of the energy scale of
inflation: the GUT scale. Other important consequences were also
discussed in the introduction.

The main issue addressed in the present article was the compatibility
of the Planck data with the BICEP2 data assuming an inflationary
prior. Several indicators have been used to quantify the tension
between these two measurements. Firstly, assuming slow-roll, we have
shown that our posterior odds measure of compatibility gives
$\calRsr\simeq 1$. This means that we are not in a position to
establish that Planck and BICEP2 are compatible at a statistically
significant level assuming a slow-roll model. But, clearly, we cannot
either prove that the two data sets are incompatible (again, assuming
slow-roll): we are precisely in a regime where one cannot
conclude. Secondly, we have also computed the $\calR$ factor for all
the \EI scenarios and shown that the undecided situation just
described is changed. We have found that the zone of strongly
compatible models contain no ``good'' Planck or BICEP2 models (\ie
``good'' models are defined to be models in the inconclusive zone with
respect to the best models of each data set alone). Moreover, all the
models for which we can be sure that Planck and BICEP2 are compatible
($\ln \calR>5$) are either strongly or moderately disfavored by Planck
(except three models that are only weakly disfavored, \ie $\bsusybil$,
$\gmssmiomB$ and $\gmssmiem$). Thirdly, for models such that
$\calR>1$, we have derived the updated value of the Bayesian
evidence. We have found that, for all the best Planck+BICEP2 models
(those which are in the inconclusive zone with respect to the best
Planck+BICEP2 model $\lfiTHREE$), we have $1<\ln \calR<5$, \ie for
none of them Planck and BICEP2 appear compatible at a strong evidence
level. Fourthly, as was established in \Refc{Martin:2013nzq}, the
Planck data favor category $1$ models, namely models with a potential
having a plateau shape (the best model was $\kmiii$ but the
inconclusive zone contained other scenarios, for instance the
Starobinsky model). However, these models are disfavored by the BICEP2
data for which the best model is $\lfiTHREE$ (a category $2$ model)
and have $\calR$-factors less than unity. Therefore, we face a
situation where Planck and BICEP2 are not strongly
compatible. Moreover, as discussed above, several hints all indicate
that the two measurements could in fact be incompatible although, in
the present situation, it is too early to make a final judgment.
\par

Another important message of this work is that, assuming BICEP2 alone
or Planck+BICEP2 when possible (\ie for $\ln \calR>1$) does not single
out a particular model, for instance $m^2\phi^2$. From a theoretical
point of view, $m^2\phi^2$ may seem a priori quite attractive.
However, given either BICEP2 or Planck+BICEP2, it is not the only winner and
other types of models are still performing as well as this simple
potential. As a consequence, in the present situation, it seems
meaningless to focus the model building efforts only on large field
models.

\par

In view of our result, the most important next step is to confirm that
the $B$-mode polarization detection by BICEP2 is truly of primordial
origin. Hopefully, this will help to resolve the tension between the
two data sets and thus their incompatibility for the Planck best
scenarios.

\par

Once done, if a the detection of a non-vanishing $r$ is confirmed, one
will have to measure the tensor spectral index $\nT$. The sign of
$\nT$ already carries very important information and has the potential
to confirm or exclude different challengers to inflation. Indeed,
inflation generically predicts a red spectrum, namely $\nT<0$, see
\Eq{eq:nst}. If one finds a blue spectrum $\nT>0$, this would
certainly be difficult (and/or contrived) to explain in this framework
and alternatives such as, for instance, string gas
cosmology~\cite{Brandenberger:2014faa}, which predicts a blue
spectrum, would be a natural solution.

\par

For a red spectrum, the next-to-next pressing question will be to
verify the simplest consistency relation of \Eq{eq:consistency},
namely~\cite{Caligiuri:2014sla, Dodelson:2014exa}
\begin{equation}
\frac{r}{\nT} \simeq -8, 
\end{equation}
which is independent of the shape of the potential (but not of the
inflationary classes of models).

\par

Only after these three steps have been completed, one would be in a
position to claim that inflation has been really seen in the sky. It
should be clear from the above considerations that this is not yet the
case.

\acknowledgments

This work is partially supported by the ESA Belgian Federal PRODEX
Grant No.~4000103071 and the Wallonia-Brussels Federation grant ARC
No.~11/15-040.

\section*{Note added}

On the same day this paper was made public, \Refc{Flauger:2014qra}
appeared and suggested that B-modes emission by polarized dust
foregrounds could have been underestimated in the BICEP2
measurements. If confirmed, such a contribution from foregrounds could
indeed help to alleviate the tension reported here between Planck and
BICEP2 for various inflationary models.

\appendix

\section{Bayesian compatibility Between Data Sets}

\begin{figure*}
\begin{center}
\includegraphics[width=0.8\wdblefig]{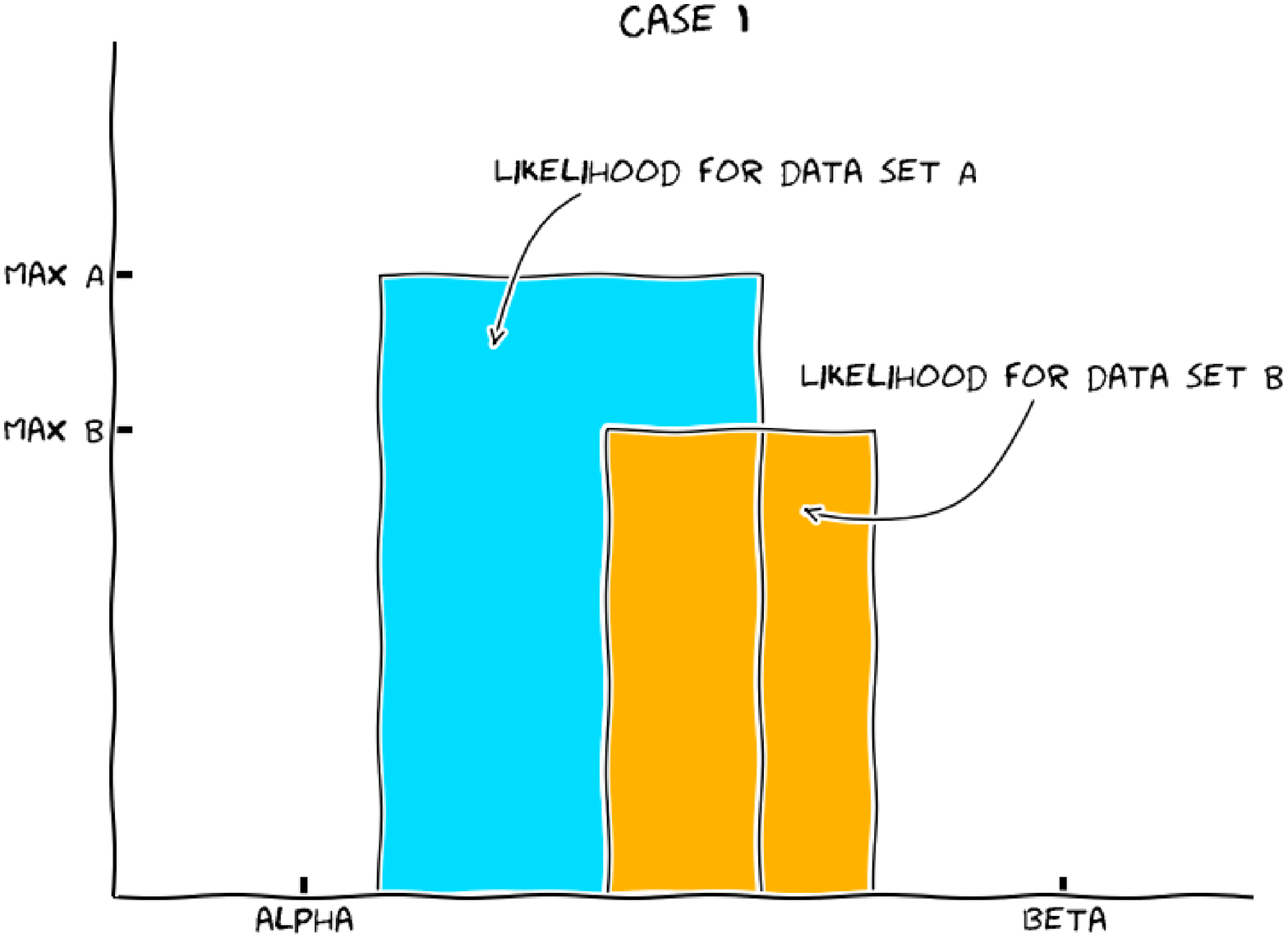}
\includegraphics[width=0.8\wdblefig]{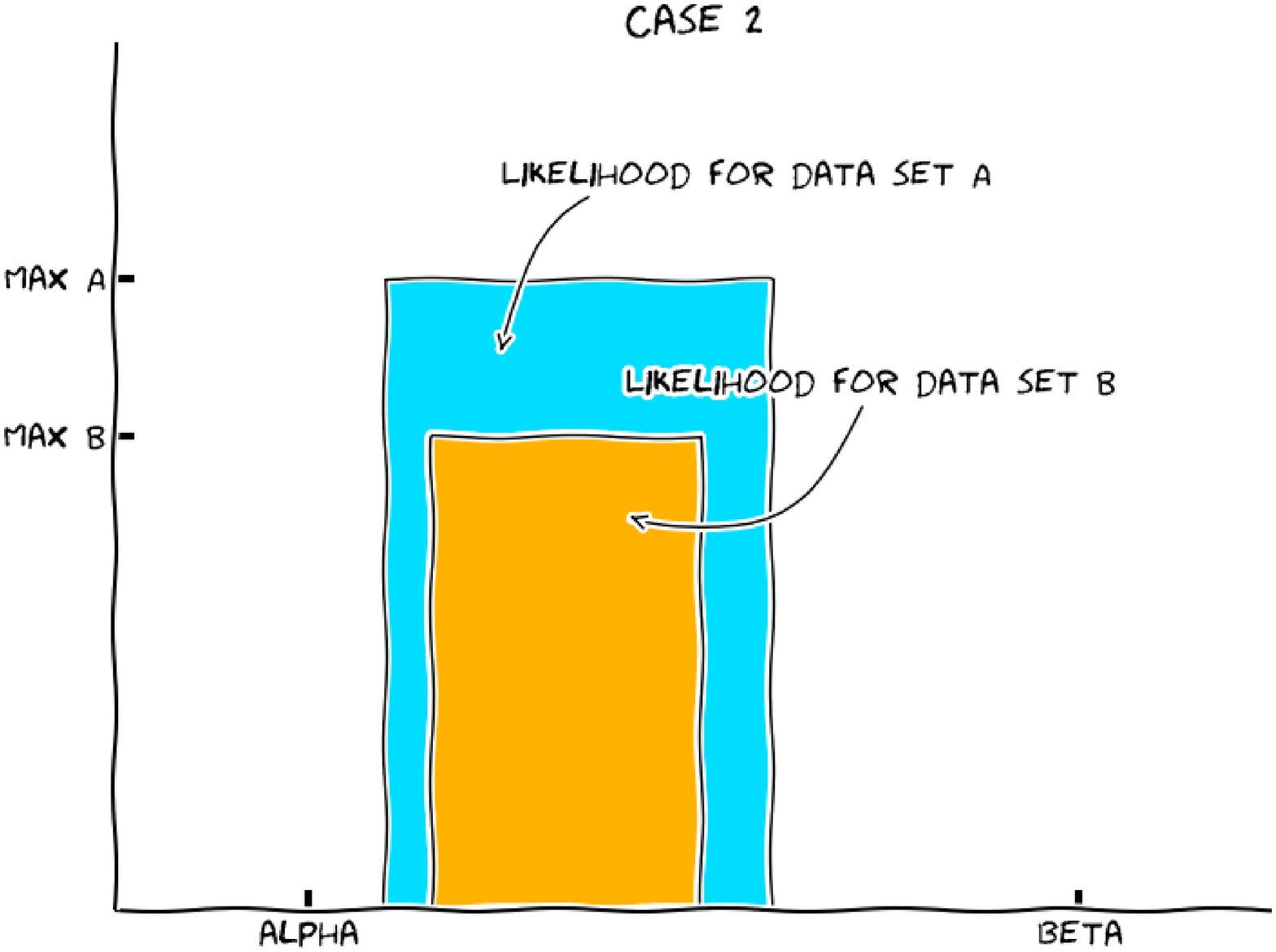}
\includegraphics[width=0.8\wdblefig]{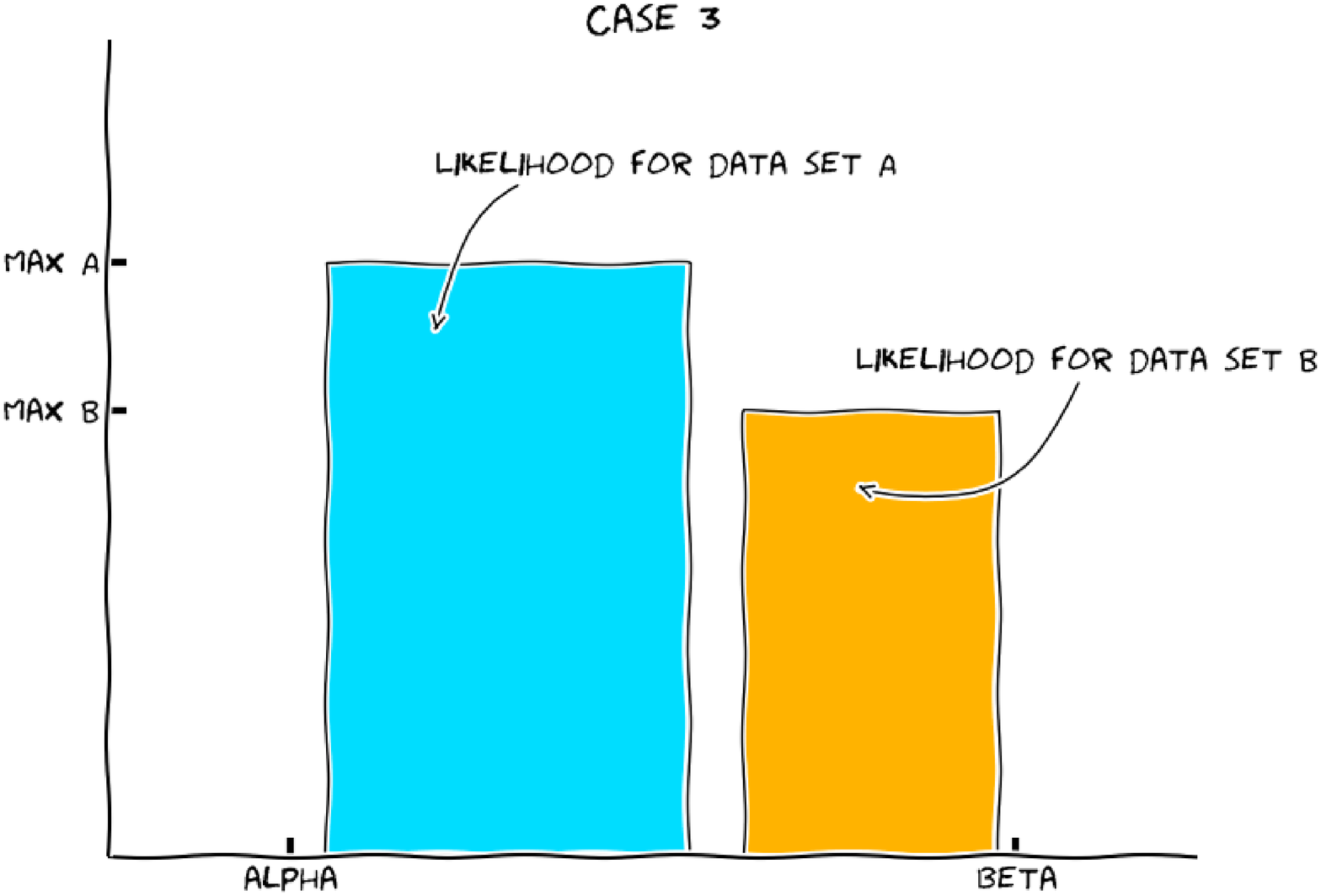}
\includegraphics[width=0.8\wdblefig]{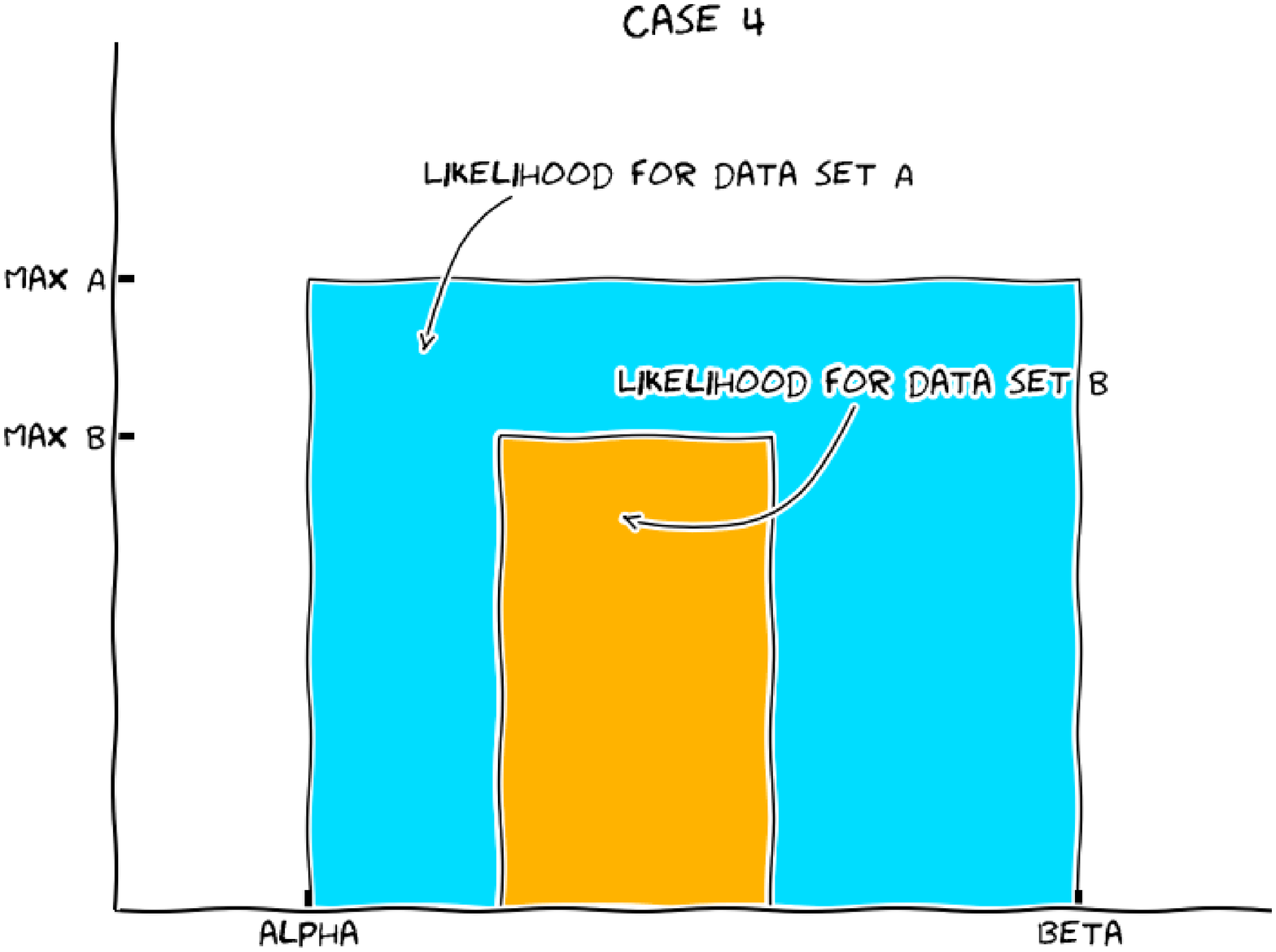}
\caption{Four proto-typical situations when combining two data sets
  $\DA$ and $\DB$. Their respective likelihoods may overlap, or not,
  within the prior volume, or not. As seen in \Fig{fig:b2sr}, Planck
  and BICEP2 under the slow-roll model prior ($\sr$) could be
  idealized as ``case 2'' for the cosmological parameters which ends
  up being constrained by BICEP2 alone ($\calR>1$), as ``case 4'' for
  those not constrained at all ($\calR=1$) and as ``case 1''
  over the $\epsstar{1}$ direction ($\calR<1$).}
\label{fig:toy}
\end{center}
\end{figure*}

In this section we illustrate how the $\calR$ factor measures the
degree of compatibility/incompatibility between two data sets given a
model $\calM$.

\subsection{Toy example}

\label{sec:toy}

We consider a toy model $\calM$ described by a single parameter
$\theta$, the prior of which is uniform in the interval
$[\alpha,\beta]$ and has a density $V_\pi^{-1}\equiv
(\beta-\alpha)^{-1}$. Let us evaluate $\calR$ associated with two data
sets $\DA$ and $\DB$ in various idealized cases as sketched in
\Fig{fig:toy}. Their respective likelihoods are assumed to be
Heaviside functions having a maximum value $\calLmax_i$ over a support
$\delta \theta_i$ ($i$ being A or B). For ``case 1'' represented in
\Fig{fig:toy}, one gets
\begin{equation}
\begin{aligned}
\calR & =\dfrac{\displaystyle \int \calL_\usssA(\theta) \calL_\usssB(\theta)
  \pi(\theta) \ud \theta}{\displaystyle \int\calL_\usssA(\theta) \pi(\theta) \ud
  \theta \int \calL_\usssB(\theta)\pi(\theta) \ud
  \theta} = \dfrac{V_\pi \delta \thetaAB}{\delta \thetaA \delta
  \thetaB} \\
& = \dfrac{\delta \thetaAB}{\min(\delta \thetaA,\delta \thetaB)} \times
\dfrac{1}{\max(\delta \thetaA, \delta \thetaB) / V_\pi}\,.
\end{aligned}
\label{eq:Rcase1}
\end{equation}
The quantity $\delta \thetaAB$ stands for the overlapping range of
$\theta$ values between the two likelihoods. We point out that the
maximum likelihood values cancel out and have no influence on
$\calR$. In the second line of \Eq{eq:Rcase1}, we have highlighted a
first factor which is always less than unity since $\delta \thetaAB
\le \min(\delta \thetaA,\delta \thetaB)$. The second term is the
inverse of factor by which the prior volume has been reduced by the
less constraining data set. Provided the less constraining data set
(i.e., the one with the largest support of the likelihood) remains
informative, namely $\max(\delta \thetaA, \delta \thetaB) < V_\pi$,
this second term in \Eq{eq:Rcase1} is always greater than unity. As
expected for a Bayesian quantity, $\calR$ measures how much the
likelihoods of the two data sets overlap balanced by how much
information has been gained with respect to the initial prior
volume. For instance, ``case 2'' in \Fig{fig:toy} yields $\delta
\thetaAB = \min(\delta \thetaA,\delta \thetaB) = \delta \thetaB$ and
$\calR = V_\pi/\delta \thetaA > 1$, so long as $\DA$ is informative
($\delta \thetaA < V_\pi$). Notice that one would get exactly the same
result for $\delta \thetaAB = \delta \thetaA = \delta \thetaB$. ``Case
4'' represents a situation in which the worse data set, here $\DA$,
becomes uninformative as the likelihood support encompasses the whole
prior volume ($\delta \thetaA = V_\pi$) and $\calR = 1$. In other
words, even though the likelihoods perfectly overlap, $\calR=1$
indicates that one cannot conclude on the compatibility of the two
data sets precisely because one of them is uninformative. Finally,
``case 3'' is the worse case scenario $\delta \thetaAB=0$ and $\calR =
0$ signaling a complete incompatibility between $\DA$ and $\DB$ under
the model $\calM$.

In view of the marginalized distributions represented in
\Fig{fig:b2sr}, the posterior of $\epsstar{1}$ exhibits a situation
typical of ``case 1''. For all the other parameters, the BICEP2
posteriors are always encompassing those associated with Planck, some
being informative and others uninformative. Therefore, some directions
in the parameter space are typical of ``case 4'' (as for instance
$\epsilon_2$ and $\epsilon_3$) while others are typical of ``case 2''
(as for instance $\thetaMC$ and $\Pstar$). As a result, one may expect
the $\calRsr$ factor between Planck and BICEP2 for the model
$\calM=\sr$ to be pushed towards unity by all the uninformative
posteriors from BICEP2, less than unity by the $\epsstar{1}$-posterior and more
than unity by the compatible posteriors; a situation more complex than
what is advocated in \Refc{Audren:2014cea}. In the following, we
provide a semi-analytic calculation confirming the numerical
calculation of \sectionc{sec:calRsr} and showing that those effects
roughly compensate to give $\calRsr$ close to unity.

\subsection{Semi-analytic approach}
\label{sec:analytical}
In the following, we split the cosmological, astrophysical and
instrumental parameters associated with the Planck likelihood into two
sets $\tetas=(\tetalcdm,\tetan)$ with
\begin{equation}
\begin{aligned}
\tetalcdm & \equiv \left(\OmegaB h^2, \OmegaCDM h^2, \tau,
100\thetaMC \right), \\
\tetan &\equiv  \left(\APSa, \APSb, \APSc, \rPSbc,\ACIBb,
\ACIBc, \rCIBbc,\right.\\
& \left. \gamCIB,\AtSZ, \AkSZ, \xitSZCIB,\ca, \cc, \betaoo \right),
\end{aligned}
\end{equation}
noticing that the BICEP2 likelihood only involves the
$\tetalcdm$ set. In order to simplify notation we denote by
$\vareps$ the set of primordial parameters $\ln(10^{10}\Pstar)$,
$\log(\epsstar{1})$, $\epsstar{2}$ and $\epsstar{3}$ and by $\Dp$ and
$\Db$ the Planck and BICEP2 data sets. From the definition of $\calRsr$
one has
\begin{equation}
\begin{aligned}
\calRsr & = \dfrac{\Evid(\Dp,\Db|\sr)}{\Evid(\Dp|\sr) \Evid(\Db|\sr)}
 \\ & = \calRref
\dfrac{\dfrac{\Evid(\Dp,\Db|\sr)}{\Evid(\Dp,\Db|\calMref)}}{\dfrac{\Evid(\Dp|\sr)}{\Evid(\Dp|\calMref)} \dfrac{\Evid(\Db|\sr)}{\Evid(\Db|\calMref)}} \,.
\end{aligned}
\label{eq:Rsrdef}
\end{equation}
Here we have introduced a reference model $\calMref$ such that the
last term in the above equation is a ratio of Bayes factors that can
be computed quickly from the effective likelihood method discussed in
\sectionc{sec:method}. The difficulty has been moved into estimating
$\calRref$, \ie the compatibility factor between Planck and BICEP2
under some reference model. However, the arbitrariness in choosing
$\calMref$ allows us to define it with a very convenient prior, namely
\begin{equation}
\pi(\vareps) = \delta(\vareps-\varepsfix),
\end{equation}
where $\varepsfix$ are some fixed values of the primordial
parameters. The evidence of $\calMref$ using Planck data alone is given by
\begin{equation}
\begin{aligned}
\Evid(\Dp|\calMref) & = \int \calLpbar(\varepsfix,\tetalcdm)
\pi(\tetalcdm) \, \ud \tetalcdm ,
\end{aligned}
\label{eq:EvidP}
\end{equation}
where we have defined
\begin{equation}
\calLpbar(\varepsfix,\tetalcdm) \equiv \int
\calLp(\varepsfix,\tetalcdm,\tetan) \pi(\tetan) \, \ud \tetan.
\end{equation}
Similarly, for Planck and BICEP2 data combined, one has
\begin{equation}
\begin{aligned}
\Evid(\Dp,\Db|\calMref)  =  \int & \calLpbar(\varepsfix,\tetalcdm) 
\calLb(\varepsfix,\tetalcdm) \\ &\times \pi(\tetalcdm) \, \ud \tetalcdm\, ,
\end{aligned}
\label{eq:product}
\end{equation}
and for the BICEP2 data alone the evidence reads
\begin{equation}
\Evid(\Db|\calMref) = \int \calLb(\varepsfix,\tetalcdm) 
\pi(\tetalcdm) \,  \ud \tetalcdm.
\label{eq:EvidB}
\end{equation}
These expressions are exact and we now make some approximations. From
the posteriors of \Fig{fig:b2sr}, one sees that, over all the
cosmological parameters $\tetalcdm$, the marginalized Planck
likelihood $\calLpbar (\varepsfix,\tetalcdm)$ is strongly peaked
inside the support of $\calLb(\varepsfix,\tetalcdm)$. Therefore,
\Eq{eq:product} can be approximated by
\begin{equation}
\begin{aligned}
\Evid(\Dp,\Db|\calMref)  & \simeq \calLb(\varepsfix,\tetalcdmMAX) \\
& \times \int \calLpbar(\varepsfix,\tetalcdm) \pi(\tetalcdm) \, \ud \tetalcdm,
\end{aligned}
\end{equation}
where $\tetalcdmMAX$ are the cosmological parameters at which
$\calLpbar$ is maximal given $\varepsfix$. From this expression,
together with \Eqs{eq:EvidP} and~(\ref{eq:EvidB}), one gets
\begin{equation}
\calRref \simeq \dfrac{\displaystyle
  \calLb(\varepsfix,\tetalcdmMAX)}{\displaystyle \int
  \calLb(\varepsfix,\tetalcdm) \pi(\tetalcdm) \, \ud
  \tetalcdm}\,,
\label{eq:Rref}
\end{equation}
which, apart from the location $\tetalcdmMAX$, depends on the BICEP2 likelihood only. The evidence appearing in
the denominator is a four-dimensional integral over $\tetalcdm$ (or
five-dimensional if one marginalizes over $\epsstar{3}$), as opposed
to a nineteen-dimensional integral for the bare Planck likelihood. In
practice, we have chosen $\varepsfix$ as the primordial parameters
associated with the best fit model of Planck and BICEP2 combined and
have evaluated \Eq{eq:Rref} using the {\MULTINEST}
algorithm~\cite{Feroz:2007kg, Feroz:2008xx}. This method yields
\begin{equation}
\ln (\calRsr) \simeq -0.01 \pm 0.3,
\label{eq:Rsranalytic}
\end{equation}
where the quoted error is a systematic estimated by performing various
nested integrations having a number of live points between $500$ and
$1000$. This value is compatible with the full numerical integration
presented in \sectionc{sec:calRsr}.

\bibliography{biblio}

\end{document}